\documentclass[acmlarge, manuscript]{acmart}

\AtBeginDocument{%
  \providecommand\BibTeX{{%
    \normalfont B\kern-0.5em{\scshape i\kern-0.25em b}\kern-0.8em\TeX}}}

\setcopyright{acmcopyright}
\copyrightyear{2020}
\acmYear{2020}
\acmJournal{DTRAP}    
\acmVolume{37}
\acmNumber{4}
\acmArticle{111}
\acmMonth{11}

\usepackage{algorithmic}
\usepackage{graphicx}
\usepackage{textcomp}
\usepackage{multicol}
\usepackage{multirow}
\usepackage{ragged2e}
\usepackage{tabularx}
\usepackage{array}
\usepackage{cancel}
\usepackage{subcaption}
\usepackage{tcolorbox}
\usepackage[htt]{hyphenat}
\usepackage{enumitem}

\begin{document}

\title{Facebook Ad Engagement in the Russian Active Measures Campaign of 2016}

\author{Mirela Silva}
\email{msilva1@ufl.edu}
\authornote{Both authors contributed equally to this research.}
\affiliation{%
    \department{Electrical \& Computer Engineering}
    \institution{University of Florida}
    \streetaddress{P.O. Box 116200}
    \city{Gainesville}
    \state{FL}
    \country{USA}
    \postcode{32611}}
    
\author{Luiz Giovanini}
\email{lfrancogiovanini@ufl.edu}
\authornotemark[1]
\affiliation{%
    \department{Electrical \& Computer Engineering}
    \institution{University of Florida}
    \streetaddress{P.O. Box 116200}
    \city{Gainesville}
    \state{FL}
    \country{USA}
    \postcode{32611}}

\author{Juliana Fernandes}
\email{juliana@jou.ufl.edu}
\affiliation{%
    \department{College of Journalism and Communications}
    \institution{University of Florida}
    \streetaddress{P.O. Box 118400}
    \city{Gainesville}
    \state{FL}
    \country{USA}
    \postcode{32611}}
    
\author{Daniela Oliveira}
\email{daniela@ece.ufl.edu}
\affiliation{%
    \department{Electrical \& Computer Engineering}
    \institution{University of Florida}
    \streetaddress{P.O. Box 116200}
    \city{Gainesville}
    \state{FL}
    \country{USA}
    \postcode{32611}}
    
\author{Catia S. Silva}
\email{catiaspsilva@ece.ufl.edu}
\affiliation{%
    \department{Electrical \& Computer Engineering}
    \institution{University of Florida}
    \streetaddress{P.O. Box 116200}
    \city{Gainesville}
    \state{FL}
    \country{USA}
    \postcode{32611}}

\renewcommand{\shortauthors}{Silva and Giovanini et al.}


\newcommand{\aka}{a.k.a.\ }
\newcommand{\etal}{\mbox{et al.\ }}
\newcommand{\ie}{\mbox{i.e.,\ }}
\newcommand{\eg}{\mbox{e.g.,\ }}
\newcommand{\etals}{\mbox{et al.'s\ }}
\newcommand{\red}[1]{\textcolor{red}{#1}}
\newcommand{\XXX}{\textcolor{red}{XXX}\ }
\newcommand{\luiz}[1]{\textcolor{blue}{#1}}
\newcommand{\totalfeat}{41}

\newcommand{\rqone}{\textbf{RQ1}}
\newcommand{\rqtwo}{\textbf{RQ2}}
\newcommand{\rqthree}{\textbf{RQ3}}
\newcommand{\rqfour}{\textbf{RQ4}}
\newcommand{\rqfive}{\textbf{RQ5}}
\newcommand{\rqsix}{\textbf{RQ6}}
\newcommand{\rqseven}{\textbf{RQ7}}

\newcommand{\rqonetext}{Is there a relationship between the ads' features and engagement?}
\newcommand{\rqtwotext}{What feature set makes a disinformation ad successful?}
\newcommand{\rqthreetext}{Given a set of the most discriminative features, how accurately can one predict engagement?}
\newcommand{\rqfourtext}{Which semantic topics best characterize the Facebook IRA ad dataset?}
\newcommand{\rqsixtext}{\ }

\newcommand{\commentario}[1]{\noindent
    \begin{tcolorbox}[colback=lime]
    \textcolor{red}{#1}
    \end{tcolorbox}
}

\newcommand{\takeaway}[1]{\noindent
    \begin{tcolorbox}[]
    {#1}
    \end{tcolorbox}
}


\begin{abstract}

This paper examines 3,517 Facebook ads created by Russia’s Internet Research Agency (IRA) between June 2015 and August 2017 in its active measures disinformation campaign targeting the 2016 U.S. general election. We aimed to unearth the relationship between ad engagement (as measured by ad clicks) and \totalfeat\ features related to ads' metadata, sociolinguistic structures, and sentiment. Our analysis was three-fold: (i) understand the relationship between engagement and features via correlation analysis; (ii) find the most relevant feature subsets to predict engagement via feature selection; and (iii) find the semantic topics that best characterize the dataset via topic modeling. We found that ad expenditure, text size, ad lifetime, and sentiment were the top features predicting users’ engagement to the ads. Additionally, positive sentiment ads were more engaging than negative ads, and sociolinguistic features (e.g., use of religion-relevant words) were identified as highly important in the makeup of an engaging ad. Linear SVM and Logistic Regression classifiers achieved the highest mean F-scores (93.6\% for both models), determining that the optimal feature subset contains 12 and 6 features, respectively. Finally, we corroborate the findings of related works that the IRA specifically targeted Americans on divisive ad topics (e.g., LGBT rights, African American reparations). 

\end{abstract}

\ccsdesc[500]{Security and privacy~Human and societal aspects of security and privacy}
\ccsdesc[500]{Security and privacy~Social network security and privacy}

\keywords{Engagement, advertisements, disinformation, Facebook, machine learning}

\maketitle

\section{Introduction}\label{sec:intro}
Disinformation is any false and deceptive content that aims to spread propaganda, promote societal division, and cast doubt in democratic processes, government institutions, and on science. 
The phenomenon is not new---disinformation has been around since humans introduced the concept of nation states~\cite{Otis2020-yj}, became notorious in Nazi Germany~\cite{bernays1928}, and was pervasively leveraged during the Cold War via the Soviet Active Measures~\cite{bittman72, Rid2020-bv}, one the the most well-documented uses of disinformation in political warfare against the U.S. and its allies. 

The account of Active Measures defectors \cite{bittman72,bittman85} sheds light on practices that remain largely the same today. There were two goals: discredit the U.S. as imperialist and permeate chaos in American and Western allies’ society. To spread disinformation, operators exploited the media's hunger for ``scoops,'' which was fed via anonymous leaks and compromised journalists. A polarized media was highly conducive to the spread of disinformation because the target wants to believe in a message that affirms their preconceived opinions. Even balloons were used to spread disinformation~\cite{Ashley_Deeks_Sabrina_McCubbin_Cody_M_Poplin2017-ut}, resulting in over 300M pamphlets littering Central Europe. Operators targeted grassroot movements to sow discord by exploiting societal vulnerabilities, such as distributing racists leaflets falsely attributed to the KKK, while simultaneously infiltrating antiracist groups~\cite{United_States_Department_of_State1986_activemeasures,Rid2020-bv}. 

The Cold War Active Measures campaigns bear a disturbing resemblance to what we are witnessing today. 
We are immersed in an environment of highly polarized, scoop-hungry media, with some outlets spreading demonstrably false information~\cite{Silverman2015-tj}.
Our society has now evolved, making room for social media to become the 21st century version of Cold War balloons spreading disinformation.
The 2019 Mueller report~\cite{Mueller2019} revealed that IRA (Internet Research Agency, associated with the Kremlin) employees travelled to the U.S. in 2014 on an intelligence-gathering mission to better understand American culture for use in social media posts. 
Arif et al.~\cite{Arif2018-vz} documented the IRA's penetration in the \#BlackLivesMatter movement, playing ``both sides'' in the discourse. 
Science continues to be leveraged as an indirect target of disinformation campaigns, inflaming the debate about climate change~\cite{Craig_Timberg2018-jo} and the coronavirus pandemic \cite{Tucker2020-fd}. 
Notably, during the 2016 U.S. presidential election, as many as 529 different rumors were spread on Twitter~\cite{Jin2017-ap}, and approximately 80,000 social media advertisements~\cite{noauthor_undated-wp} were identified by the United States House of Representatives Permanent Select Committee on Intelligence (HPSCI) as disinformation advertisements released by Russian actors with the intent of interfering with the 2016 presidential campaign and sow division in American society by exploring issues such as race (Black Lives Matter advocacy), 2nd amendment rights, and immigration.

McFaul~\cite{McFaul2018-np} gives evidence that Soviet Active Measures never stopped: the U.S. went from a Cold War with the Soviet Union to a Hot Peace with Russia.  
The key difference between disinformation now and in the last century is that the Internet and social media platforms have amplified disinformation's scope, speed, and detrimental effects. While in the past campaigns were expensive, long, and ``manual'' (e.g., flyers disseminated from the sky via balloons~\cite{bittman72, Rid2020-bv}, spreading disinformation today is arguably cheaper, faster (click of a button), and executed remotely, complicating attribution.
Tackling disinformation is difficult because: (i)  spreading it is not illegal in the United States, (ii) solutions cannot infringe freedom of speech, (iii) dissemination speed and scale can render fact-checkers quickly outdated, and (iv) the combination of truth with falsehoods exacerbates human confusion and challenges automatic detection.  

What makes the disinformation campaigns surrounding the U.S. presidential election remarkable is that it was one of the best well-documented Active Measures Russia conducted against the US since the Cold War. Analysis on this campaign is essential for defenses against future campaigns because Russia's Active Measures will not stop. 
In fact, the Senate Select Committee on Intelligence report on Active Measures on social media~\cite{Select_committee_on_intelligence2019-fb} highlights that IRA activity on social media did not cease, but rather increased after Election Day 2016, as if the results emboldened the Russia government~\cite{Hindman2018-if, Knight_Foundation2018-vk}. 
Moreover, reports have shown that foreign states such as Russia, China, and Iran targeted the Donald Trump and Joe Biden 2020 election campaigns in the U.S., using similar techniques as those employed by the IRA in 2016~\cite{BBC_News2020-ve}.

In this paper, we analyze a dataset~\cite{noauthor_undated-wp} made available by the U.S. House of Representatives Permanent Select Committee on Intelligence containing 3,517 Facebook ads created by the Russian Internet Research Agency (IRA) from June 2015 to August 2017. 
We hypothesize that the number of clicks reflects an ad's pertinence and users' engagement; therefore, we opted to use \textbf{ad clicks as our engagement metric}.
In predicting ad engagement (measured by ad clicks), we identified four broad categories of features with potential predictive value: 
(1) the ad's metadata (e.g., lifetime, expenditure);
(2) the size of the ad's text (character and word count);
(3) the ad text's sociolinguistic features (e.g., authenticity and emotional tone); and
(4) the ad text's subjectivity (e.g., objective vs. subjective) and sentiment (e.g., positive vs. negative).
We therefore aimed to investigate:

\begin{itemize}
    \item \rqone: \rqonetext
    \item \rqtwo: \rqtwotext
    \item \rqthree: \rqthreetext
    \item \rqfour: \rqfourtext

\end{itemize}

In quantifying our empirical investigation, we implemented several correlation analysis methods, as well as machine learning analysis for topic modeling and feature selection. 
This paper confirms and builds on major findings from similar works, such as~\cite{DiResta_IRA_report, Mueller2019, howard18}. 
We confirm that communities (e.g., African Americans, Republicans, LGBT) were specifically targeted by the IRA to sow dissent within American society, and several communities experienced an increase in engagement with the Russian ads in our dataset during key moments of the 2016 presidential election (e.g., during President Trump's office takeover).
However, in contrast to DiResta et al.~\cite{DiResta_IRA_report}, who performed a qualitative analysis of the ads, we do so through a quantitative methodology, combining statistics and multi-methods machine learning focused on engagement.
Many aspects of our results corroborate prior works~\cite{DiResta_IRA_report, alvarez2020, howard18}, but we also go further than prior works to show that high engagement ads were more positive in terms of sentiment, more informal and personal, and shorter in text size than standard engagement ads. 
Finally, we find that ad expenditure was ranked as the most important feature for predicting high engagement by six machine learning models, and that sociolinguistic features of the ad (e.g., the presence of words associated with religion) made up the top 5 features for predicting high engagement for the majority of the learning models.

This paper is organized as follows.
Section~\ref{sec:related_works} reviews prior works analyzing the Russian Active Measures disinformation campaign related to the U.S. presidential election of 2016.
Section~\ref{sec:methodology} describes the methodology of our analysis. 
Section~\ref{sec:results} presents our correlation analyses and machine learning results. 
Section~\ref{sec:discussion} discusses our study's findings, limitations, and future work directions. 
Section~\ref{sec:conclusion} concludes the paper.


\section{Related Works}\label{sec:related_works}

In this section, we focus on prior work intersecting the topic of the present paper, in particular prior analyses of Russia’s great active measures campaign of 2016 and disinformation spreading.


Investigations and reports on Russian efforts to influence the 2016 U.S. elections emerged as early as mid-2016 via the FBI Crossfire Hurricane investigation and  after Congress members had access to classified intelligence~\cite{Miller2016-du}. 
After the election, in early 2017, the Office of the Director of National Intelligence released an assessment of the Russian influence and disinformation campaign \cite{Office_of_the_Director_of_National_Intelligence2017-tx}, for the first time acknowledging its similarities to the Soviet Active Measures campaigns that targeted the U.S. during the Cold War \cite{bittman72,bittman85,Perkins2018-pu,Rid2020-bv}. 
The report highlighted a perceived change in Russia intelligence efforts, which since the Cold War, have been primarily focused on foreign intelligence collection. 
For decades, Russian and Soviet intelligence services have sought to collect insider information to allow the Kremlin with a better understanding of U.S. priorities and foreign policy.
However, the Intelligence Community had uncovered that Vladimir Putin had ordered an influence campaign using social media to hurt Clinton's electoral chances and undermine public faith in the U.S.'s democratic process.  
Next, Congress sought the aid of experts and social media companies in facilitating its public hearings and investigations. 
Following the firing of FBI Director James Comey, a Special Counsel was formed and represented another line of investigation on Russian active measures campaign. 
In September 2017, the media started reporting~\cite{Strohm2017-hx} that the Mueller probe was focused on the use of social media as the main tool for the active measures campaign. 
This prompted social media companies to conduct internal audits, which led to a dataset of tweets, Facebook ads and posts, and YouTube videos being released to the House Permanent Select Committee on Intelligence.

The Senate Select Committee on Intelligence undertook a study of these events and sought the input of two main Technical Advisory Groups (TAG) to analyze the dataset provided to the Committee by the social media companies~\cite{howard17,howard18,DiResta_IRA_report}. 
Both groups analyzed thousands of ads, pages, tweets, and posts that social media companies independently identified through audits pertaining to the Internet Intelligence Agency's (IRA) active measures campaigns; the analyses focused on qualitative and quantitative aspects of the dataset. 
Both reports, released to the public in late 2018, reached similar conclusions, corroborated in early 2019 by the Mueller report~\cite{Mueller2019}. 
The IRA, supported by the Kremlin, conducted a major active measures campaign in the years preceding the 2016 presidential election campaign, with their social media stimuli reaching millions of American citizens. 
They sought two main goals: 
(1) influence the 2016 U.S. presidential election by harming Hillary R. Clinton's chances of success while supporting then-candidate Donald J. Trump, and 
(2) sow discord in American politics and society, especially on race issues by heavily targeting the African American population while playing both sides of the political discourse~(also corroborated by independent work from Arif et al.~\cite{Arif2018-vz}). 
The group led by John Kelly~\cite{howard18} also stressed the key role played by Twitter bots in amplifying propaganda, in agreement with prior research by Bessi and Ferrara et al.~\cite{bessi16,ferrara16}. 

The intelligence reports and independent researchers also analyzed the IRA Facebook paid advertisements (ads) from qualitative and quantitative perspectives. 
In particular, the House Permanent Select Committee on Intelligence released 3,517 Facebook ads associated with the IRA in 2018. 
Although the ads were not the bulk of the IRA's activity in social media, the use of advertising was consistent with IRA’s modus operandi~\cite{Select_committee_on_intelligence2019-fb}: divisive subjects related to race, police brutality, Second Amendment rights, patriotism, LBGT rights, and immigration \cite{Kim2018-cv}. 
In a U.S. census-representative survey, Ribeiro et al.~\cite{Ribeiro2019-ns} found that people from different socially salient groups react differently to the content of the IRA's Facebook ads, further positing that Facebook's ad API facilitated this divisive targeting. 
Indeed, Facebook estimates that 11.4M Americans saw at least one of the ads ultimately determined to have been purchased by the IRA \cite{Select_committee_on_intelligence2019-fb}.
The work closest to ours is by Alvarez et al.~\cite{alvarez2020}, who performed sentiment analysis on this same Facebook ads dataset to correlate positive and negative emotions with engagement and discover how the valence of emotions changed over time. 
The analyses found negative sentiment was more prevalent before the elections and positive sentiments after the election. 
Through the use of the versatile Maximal Correlation analysis, we confirm that positive sentiment ads were correlated with high engagement. 

While previous works have focused on generation, measurement, and content of propaganda, the goal of this research was to assess in-depth the effectiveness (i.e., engagement) of such tactics. 
Thus, this paper expands these prior works by focusing only on the Facebook ads to find correlations between user engagement and \totalfeat\ features (e.g., sentiment, sociolinguistic features of the ad's text).
We also compared six machine learning models for feature selection, to further analyze which ad features were most important for engagement.
We further leveraged Latent Dirichlet Allocation (LDA) to detect, in an unsupervised fashion, eight major topics/groups (e.g., justice and African American, LGBT rights) weaponized in the ads; this confirms Howard et al.'s~\cite{howard18} analysis, wherein 20 clusters of audiences/groups (e.g., African American politics and culture, black identity and nationalism, LGBT rights and social liberalism) were identified using modularity to find community structures in networks.

A deep understanding of engagement is imperative to effectively measure disinformation, yet current researchers argue that measuring disinformation is likely impossible. 
For example, the TAG group led by DiResta et al.~\cite{DiResta_IRA_report} argued that determining whether the IRA's disinformation campaigns indeed affected the 2016 presidential election is impossible. 
Rid~\cite{Rid2020-bv} similarly argued that it is unlikely that the Russian trolls convinced a significant number of  American voters to change their minds because  the volume of IRA activity was lower than reported: only 8.4\% of IRA activity was election-related~\cite{twitter18} and the discourse happened in echo-chambers where people already had their minds set.  
However, former Soviet disinformation defectors such as Ladislav Bittman beg to differ~\cite{bittman72,bittman85}---disinformation can indeed be measured. 
In his account of Soviet disinformation tactics, Bittman discussed the two ways by which the KGB measured the success of disinformation campaigns. 
The first was through the attention (i.e., engagement) that the message was drawing outside the Soviet bloc, e.g., the amount of public discussion generated by the message and the tone of the political discourse on the issue. 
In the 21st century, this metric is what online platforms call \textit{engagement}: a function of the number of article/post views, likes, retweets, shares, mentions, etc. 
Bittman stressed the cult of the published word: the number of words used by the mass media of the enemy or victim is more important than a careful evaluation of the operation results.  
Less attention is paid to whether the words had the desired effect. 
The second metric to measure disinformation was determining whether the message forced the target country to make any political changes that could directly or indirectly benefit the Soviet Union. 
In the 21st century, the election of President Donald J. Trump could be a political change that benefited Russia's political interests, as the U.S. intelligence community confirmed~\cite{Mueller2019}.
 
According to Bittman, the Soviet Union knew that it was unlikely that a single disinformation campaign would tip the balance of power. 
However, disinformation operatives like himself believed that mass production of propaganda and disinformation over several decades would have a significant effect. 
The same rationale applies today: one tweet or Facebook post may not tip the balance, however, several months of posts on a disinformation narrative (e.g., questioning the integrity of a presidential election) might cause irreparable harm to a democracy. 
Our paper and analyses provide in-depth insights on engagement as a key metric of disinformation impact.

\section{Dataset \& Feature Extraction}\label{sec:methodology}

This section describes the dataset used in our analyses along with the steps taken for data cleaning and feature extraction.


\subsection{Dataset Description \& Filtering}
We leveraged a dataset of 3,517 Facebook ads created by the Russian Internet Research Agency (IRA) and made publicly available to the U.S. House of Representatives Permanent Select Committee on Intelligence~\cite{noauthor_undated-wp} by Facebook after internal audits. 
Estimated to have been exposed to over 126M Americans between June 2015 and August 2017, these ads were a small representative sample of over 80,000 organic content identified by the Committee. Of the 3,517 ads, 3,290 contained text entry; the remaining 227 ads were purged from the dataset, as we were interested in performing sentiment analysis and topic modeling based on the ads' text. 
Next, we discarded four ads that did not contain a numerical value for the number of ad clicks (our criteria for measuring engagement). 
Therefore, \textbf{our final dataset contained 3,286 Facebook ads} created by the IRA. 
Most of these ads (52.8\%) were posted in 2016 (the U.S. election year), followed by 29.2\% in 2017, and the remaining 18.0\% in 2015.

\subsection{Feature Extraction}
For each of the 3,286 ads, we extracted a total of \totalfeat\ features (see Table~\ref{tab:dataset_info}) that can be summarized into four main categories: 
(i) \textit{ad metadata features}, extracted from the metadata already contained in the dataset (e.g., \# of ad clicks and impressions); 
(ii) \textit{text size features}, related to the size of the text itself (e.g., word count); 
(iii) \textit{sentiment \& subjectivity features}, describing both valence (positive vs. negative) and salience (low to high arousal) of sentiment in the ad's text; and 
(iv) \textit{sociolinguistic features}, related to emotions, mood, and cognition present in the ad's text based on word counts (e.g., the words ``crying," ``grief," and ``sad" are counted as expressing sadness).

\begin{table}[h]
\caption{Summary of all features for each engagement group. For a detailed list and explanation of LIWC's sociolinguistic features, see~\cite{pennebaker2015development}.}
\label{tab:dataset_info}
\resizebox{\textwidth}{!}{%
\begin{tabular}{|c|c|ccc|ccc|}
\hline
\multirow{2}{*}{\textbf{Feature Category}} & \multirow{2}{*}{\textbf{Feature}} & \multicolumn{3}{c|}{\textbf{\begin{tabular}[c]{@{}c@{}}Standard Engagement\\ (Ad Clicks $<2,188, N = 2854$)\end{tabular}}} & \multicolumn{3}{c|}{\textbf{\begin{tabular}[c]{@{}c@{}}High Engagement\\ (Ad Clicks $\geq 2,188, N = 432$)\end{tabular}}} \\
 &  & \textbf{Min} & \textbf{Mean} & \textbf{Max} & \textbf{Min} & \textbf{Mean} & \textbf{Max} \\ \hline
\multirow{4}{*}{\textit{Ad Metadata}} & \textbf{Ad Clicks} & 0 & 297.81 & 2182.00 & 2214.00 & 6248 & 73063.00 \\
 & \textbf{Ad Impressions} & 0 & 3715.92 & 165121.00 & 8429.00 & 65223 & 1334544.00 \\
 & \textbf{Ad Lifetime} (hours) & 0 & 125.60 & 6722.42 & 16.63 & 59 & 1200.34 \\
 & \textbf{Ad Spend} (RUB) & 0 & 917.22 & 27500.00 & 100.00 & 7311 & 331675.75 \\ \hline
\multirow{2}{*}{\textit{Text Size}} & \textbf{Character Count} & 7 & 270.20 & 2716.00 & 6.00 & 163 & 1641.00 \\
 & \textbf{Word Count} & 0 & 44.72 & 437.00 & 1.00 & 26 & 274.00 \\ \hline
\multirow{14}{*}{\textit{Sentiment \& Subjectivity}} & \textbf{NLTK VADER Compound Score} & -1 & 0.08 & 1.00 & -0.99 & 0.17 & 0.97 \\
 & NLTK Negative Sentiment Only & -1 & -0.24 & 0.00 & -0.99 & -0.12 & 0.00 \\
 & NLTK Positive Sentiment Only & 0 & 0.32 & 1.00 & 0.00 & 0.30 & 0.97 \\
 & NLTK Neutral Sentiment Only (Binary) & 0 & 0.15 & 1.00 & 0.00 & 0.28 & 1.00 \\ \cline{2-8} 
 & \textbf{TextBlob Sentiment Polarity} & -1 & 0.10 & 1.00 & -0.80 & 0.11 & 1.00 \\
 & TextBlob Negative Sentiment Only & -1 & -0.05 & 0.00 & -0.80 & -0.04 & 0.00 \\
 & TextBlob Positive Sentiment Only & 0 & 0.15 & 1.00 & 0.00 & 0.15 & 1.00 \\
 & TextBlob Neutral Sentiment Only (Binary) & 0 & 0.25 & 1.00 & 0.00 & 0.38 & 1.00 \\ \cline{2-8} 
 & \textbf{Flair Sentiment} & 0 & 0.59 & 1.00 & 0.00 & 0.70 & 1.00 \\
 & Flair Negative Sentiment Only (Binary) & 0 & 0.41 & 1.00 & 0.00 & 0.30 & 1.00 \\
 & Flair Positive Sentiment Only (Binary) & 0 & 0.59 & 1.00 & 0.00 & 0.70 & 1.00 \\ \cline{2-8} 
 & \textbf{TextBlob Subjectivity} & 0 & 0.39 & 1.00 & 0.00 & 0.35 & 1.00 \\
 & TextBlob Subjective Scores Only & 0 & 0.23 & 1.00 & 0.00 & 0.23 & 1.00 \\
 & TextBlob Objective Scores Only & 0 & 0.16 & 0.50 & 0.00 & 0.12 & 0.50 \\ \hline
\multirow{21}{*}{\textit{Sociolinguistic (LIWC)}} & \textbf{Analytical Thinking} & 0 & 70.27 & 99.00 & 1.00 & 60.10 & 99.00 \\
 & \textbf{Authentic} & 0 & 27.24 & 99.00 & 1.00 & 23.80 & 99.00 \\
 & \textbf{Clout} & 0 & 75.91 & 99.00 & 1.00 & 74.69 & 99.00 \\
 & \textbf{Emotional Tone} & 0 & 47.50 & 99.00 & 1.00 & 45.85 & 99.00 \\\cline{2-8} 
 & \textbf{Affective Processes} & 0 & 7.80 & 100.00 & 0.00 & 7.81 & 100.00 \\
 & \textbf{All Punctuation} & 0 & 22.10 & 180.00 & 0.00 & 23.19 & 150.00 \\
 & \textbf{Biological Processes} & 0 & 2.06 & 33.33 & 0.00 & 1.61 & 25.00 \\
 & \textbf{Cognitive Processes} & 0 & 8.40 & 100.00 & 0.00 & 10.09 & 71.43 \\
 & \textbf{Death} & 0 & 0.41 & 50.00 & 0.00 & 0.47 & 50.00 \\
 & \textbf{Drives} & 0 & 14.31 & 100.00 & 0.00 & 13.58 & 100.00 \\
 & \textbf{Future Focus} & 0 & 0.67 & 50.00 & 0.00 & 0.82 & 20.00 \\
 & \textbf{Past Focus} & 0 & 1.96 & 50.00 & 0.00 & 2.32 & 50.00 \\
 & \textbf{Present Focus} & 0 & 10.59 & 100.00 & 0.00 & 11.18 & 57.14 \\
 & \textbf{Home} & 0 & 0.34 & 28.57 & 0.00 & 0.22 & 25.00 \\
 & \textbf{Leisure} & 0 & 1.51 & 33.33 & 0.00 & 1.21 & 33.33 \\
 & \textbf{Money} & 0 & 1.04 & 20.00 & 0.00 & 0.52 & 20.00 \\
 & \textbf{Perceptual Processes} & 0 & 4.59 & 100.00 & 0.00 & 4.85 & 50.00 \\
 & \textbf{Relativity} & 0 & 11.35 & 66.67 & 0.00 & 9.73 & 100.00 \\
 & \textbf{Religion} & 0 & 0.98 & 66.67 & 0.00 & 0.64 & 33.33 \\
 & \textbf{Social Processes} & 0 & 13.13 & 80.00 & 0.00 & 14.84 & 66.67 \\
 & \textbf{Work} & 0 & 2.72 & 50.00 & 0.00 & 2.47 & 100.00 \\ \hline
\end{tabular}%
}
\end{table}

\subsubsection{Ad Metadata \& Engagement}
The dataset consisted of one PDF file for each Facebook ad. A typical PDF datum was composed of 2 pages, where the first page contained ad metadata (e.g., the textual content of the ad, the link to the ad) and the second page contained a screenshot of the ad as seen by Facebook users (see Fig.~\ref{fig:visceral_images}for examples). We used the PyPDF2 Python library~\cite{pypdf2-jt} to automatically extract the following metadata features from each ad: 


\begin{itemize}
    \item \textbf{Ad Impressions:} the number of users who viewed the ad.
    \item \textbf{Ad Clicks:} the number of users who clicked on the ad.
    \item \textbf{Ad Spend:} the amount of money (in RUB) spent on the ad.
    \item \textbf{Ad Lifetime:} the ad's creation and end dates (in hours).
\end{itemize}


Engagement includes all actions that users take in reaction to an advertisement, such as viewing, clicking, liking, commenting, and sharing. Because the metadata made available for the dataset only captures two of the aforementioned actions (ad impressions and clicks), we opted to use the feature Ad Clicks as our metric for ad engagement. Ad Clicks is a good measure because it indicates how many users actually engaged with the advertisement, i.e., took action by clicking on the ad after exposure.
We opted to disregard Ad Impressions in our analyses as it was highly correlational with Ad Clicks (see Sec.~\ref{sec:corr_analysis} and \ref{sec:limitations}).

\subsubsection{Text Size}
We summarized the length of the ad's text using a total of two features: \textbf{character count} and \textbf{word count}.

\subsubsection{Sentiment \& Subjectivity Analysis}\label{sec:methodology_sentiment}
We leveraged three sentiment analysis packages:

\begin{itemize}
    \item\textbf{VADER}~\cite{gilbert2014vader}: a rule-based NLP library; outputs a uni-dimensional and normalized \textit{compound score} that ranges from $-1.0$ (negative) to $1.0$ (positive), where scores between $-0.05$ and $0.05$ are considered neutral sentiment.
    \item\textbf{TextBlob}~\cite{loria2014TextBlob}: a rule-based NLP library; outputs a  \textit{polarity} (sentiment) score that ranges from $-1.0$ (negative) to $1.0$ (positive) sentiment, as well as a \textit{subjectivity} score ranging from $0.0$ (objective) to $1.0$ (subjective). 
    \item\textbf{Flair}~\cite{akbik2018coling}: an embedding-based framework built on PyTorch; Flair's pre-trained sentiment model outputs labels of either \texttt{POSITIVE} or \texttt{NEGATIVE} sentiment.
\end{itemize}

We validated these sentiment analysis packages using the average F-score as the performance metric with a dataset containing 50K highly-polarized movie reviews from IMDB~\cite{maas-EtAl:2011:ACL-HLT2011}.
In this dataset, 25K reviews were labeled positive and 25K negative; Flair greatly outperformed both VADER and TextBlob (89.5\% vs. 69.0\% vs. 66.5\%, respectively). 
Nonetheless, we opted to use all 14 sentiment and subjectivity features in our analyses as listed in Table~\ref{tab:dataset_info}.

\subsubsection{Sociolinguistic Features}
To extract sociolinguistic features, we leveraged LIWC2015~\cite{pennebaker2015development}, a text analysis tool that reflects a text's emotions, thinking styles, social concerns, and grammar (e.g., parts of speech) based on word counts. A total of 21 LIWC features were extracted:

\begin{itemize}
    \item Four \textbf{summary variables}: analytical thinking (formal, logical, and hierarchical thinking vs. informal, personal, here-and-now, and narrative thinking), clout (expertise and confidence vs. tentative, humble, or anxious), authenticity (honest, personal, and disclosing text vs. guarded or distanced), and emotional tone (positive, upbeat style vs. anxiety, sadness, or hostility; values around 50 suggesting neutrality or ambivalence), each measured on a 100-point scale. 
    
    \item Seventeen other \textbf{LIWC categories}, most of which are related to psychological processes. Each of these features was measured as percentage of words (e.g., ``affective process" of 10 means that 10\% of all words of the ad's text were related to emotions, such as ``happy'' and ``cried"). See Table~\ref{tab:liwc_features} for more detailed examples of these 17 features:
    affective processes (e.g., positive emotions, anxiety, anger), social processes (e.g., family, friends), cognitive processes (e.g., insight, certainty, discrepancy), perceptual processes (e.g., see, hear, feel), biological processes (e.g., body, health, sexual), drives (e.g., affiliation, power, reward), time orientations (past/present/future focus), relativity (e.g., motion, space, time), personal concerns (e.g., work, leisure activities), and punctuation (e.g., periods, commas). 
\end{itemize}


\begin{table}[]
\centering
\caption{Example keywords for select LIWC's sociolinguistic features~\cite{pennebaker2015development}.}
\resizebox{0.70\textwidth}{!}{%
\label{tab:liwc_features}
\begin{tabular}{@{}lcll@{}}
\toprule
\multicolumn{1}{c}{\textbf{Category}} & \textbf{\# of words in category} & \multicolumn{1}{c}{\textbf{Subcategories}} & \multicolumn{1}{c}{\textbf{Example of words}} \\ \midrule
\multirow{2}{*}{Affective processes}  & \multirow{2}{*}{1393}            & Positive emotion                           & love, nice, sweet                             \\ \cmidrule(l){3-4} 
                                      &                                  & Negative emotion                           & hurt, ugly, nasty                             \\ \midrule
\multirow{4}{*}{Social processes}     & \multirow{4}{*}{756}             & Family                                     & daughter, dad, aunt                           \\ \cmidrule(l){3-4} 
                                      &                                  & Friends                                    & buddy, neighbor                               \\ \cmidrule(l){3-4} 
                                      &                                  & Female references                          & girl, her, mom                                \\ \cmidrule(l){3-4} 
                                      &                                  & Male references                            & boy, his, dad                                 \\ \midrule
\multirow{6}{*}{Cognitive processes}  & \multirow{6}{*}{797}             & Insight                                    & think, know                                   \\ \cmidrule(l){3-4} 
                                      &                                  & Causation                                  & because, effect                               \\ \cmidrule(l){3-4} 
                                      &                                  & Discrepancy                                & should, would                                 \\ \cmidrule(l){3-4} 
                                      &                                  & Tentative                                  & maybe, perhaps                                \\ \cmidrule(l){3-4} 
                                      &                                  & Certainty                                  & always, never                                 \\ \cmidrule(l){3-4} 
                                      &                                  & Differentiation                            & hasn't, but, else                             \\ \midrule
\multirow{3}{*}{Perceptual processes} & \multirow{3}{*}{436}             & See                                        & view, saw, seen                               \\ \cmidrule(l){3-4} 
                                      &                                  & Hear                                       & listen, hearing                               \\ \cmidrule(l){3-4} 
                                      &                                  & Feel                                       & feels, touch                                  \\ \midrule
\multirow{4}{*}{Biological processes} & \multirow{4}{*}{748}             & Body                                       & cheek, hands, spit                            \\ \cmidrule(l){3-4} 
                                      &                                  & Health                                     & clinic, flu, pill                             \\ \cmidrule(l){3-4} 
                                      &                                  & Sexual                                     & horny, love, incest                           \\ \cmidrule(l){3-4} 
                                      &                                  & Ingestion                                  & dish, eat, pizza                              \\ \midrule
\multirow{5}{*}{Drives}               & \multirow{5}{*}{1103}            & Affiliation                                & ally, friend, social                          \\ \cmidrule(l){3-4} 
                                      &                                  & Achievement                                & win, success, better                          \\ \cmidrule(l){3-4} 
                                      &                                  & Power                                      & superior, bully                               \\ \cmidrule(l){3-4} 
                                      &                                  & Reward                                     & take, prize, benefit                          \\ \cmidrule(l){3-4} 
                                      &                                  & Risk                                       & danger, doubt                                 \\ \midrule
\multirow{3}{*}{Time orientations}    & \multirow{3}{*}{862}             & Past focus                                 & ago, did, talked                              \\ \cmidrule(l){3-4} 
                                      &                                  & Present focus                              & today, is, now                                \\ \cmidrule(l){3-4} 
                                      &                                  & Future focus                               & may, will, soon                               \\ \midrule
\multirow{3}{*}{Relativity}           & \multirow{3}{*}{974}             & Motion                                     & arrive, car, go                               \\ \cmidrule(l){3-4} 
                                      &                                  & Space                                      & down, in, thin                                \\ \cmidrule(l){3-4} 
                                      &                                  & Time                                       & end, until, season                            \\ \midrule
\multirow{6}{*}{Personal concerns}    & \multirow{6}{*}{1314}            & Work                                       & job, majors, xerox                            \\ \cmidrule(l){3-4} 
                                      &                                  & Leisure                                    & cook, chat, movie                             \\ \cmidrule(l){3-4} 
                                      &                                  & Home                                       & kitchen, landlord                             \\ \cmidrule(l){3-4} 
                                      &                                  & Money                                      & audit, cash, owe                              \\ \cmidrule(l){3-4} 
                                      &                                  & Religion                                   & altar, church                                 \\ \cmidrule(l){3-4} 
                                      &                                  & Death                                      & bury, coffin, kill                            \\ \bottomrule
\end{tabular}
}
\end{table}

\section{Data Analysis \& Results}\label{sec:results}

This section describes the diverse set of analyses conducted to answer our research questions, along with our results identifying several features that prompt users' engagement with disinformation. 
Specifically, we performed the following analyses:

\begin{itemize}
    \item A correlation analysis described in Sec.~\ref{sec:corr_analysis} to answer \rqone~("\rqonetext").
    
    \item A feature selection analysis detailed in Sec.~\ref{sec:results_feature_selection} to answer \rqtwo~("\rqtwotext") and \rqthree~("\rqthreetext").
    
    \item A topic modeling analysis discussed in Sub.~\ref{sec:topic_modeling} to answer \rqfour~("\rqfourtext").
    
\end{itemize}


\subsection{Correlation Analysis (\rqone: Relationship Between Features and Engagement)}\label{sec:corr_analysis}
We opted to separate the dataset into a standard group and an outlier group to better understand how low engagement vs. high engagement vary as a function of ad features. Using the 1.5xIQR rule (i.e., values above $Q3 + 1.5\times IQR$), we identified 432 upper outliers based on Ad Clicks.
Therefore, ads with $< 2,188$ clicks $(n = 2,854, 86.9\%)$ were assigned to the \texttt{Standard Engagement} group (subscript \textit{stand}) and those with $\geq 2,188$ clicks $(n = 432, 13.1\%)$ were assigned to the \texttt{High Engagement} group (subscript \textit{high}).

Before we could perform statistical analyses on the extracted features, we used the Shapiro-Wilk to test for normality and found that none of the continuous metadata features (Ad Clicks, Impressions, Spend, and Lifetime) was normally distributed ($p<.001$ for all variables) based on a 1\% significance level. Prior to calculating Pearson's and Spearman's Rank Correlation Coefficients, we normalized the continuous metadata features using the Yao-Johnson power transformation (as it allows for zero and negative values) because these features exhibited a heavy positive skew (i.e., the Fisher-Pearson coefficient of skewness was $>>1$). This removed the skewness from the ad metadata features.

We used Pearson's ($r$) and Spearman's Rank Correlation ($\rho$) tests to find the linear and monotonic correlations, respectively, between Ad Clicks (our engagement metric) and all other extracted features (see Table~\ref{tab:correlation_results}). 
The coefficients for both tests range from -1 to 1, where negative values suggest a negative correlation between two variables (i.e., one variable decreases as the other variable increases and vice-versa) while positive values suggest a positive correlation (i.e., one variable increases/decreases as the other variable increases/decreases). 
Our interpretations regarding the strength of the correlations were based on the following intervals~\cite{Corder2011-eg}: 
$r,\rho=0$ denotes no correlation (i.e., the two variables are completely independent of each other);
$0<|r,\rho|<0.1$ denotes a trivial correlation; 
$0.1\leq |r,\rho|<0.3$ a weak/small correlation; 
$0.3\leq |r,\rho|<0.5$ a moderate/medium correlation;
$0.5\leq |r,\rho|<1$ a strong/large correlation; and 
$|r,\rho|=1$ a perfect correlation.

\begin{table}[h]
\caption{Correlation analyses for Ad Clicks (dependent variable) vs. features. Note that the range for Maximal Correlation (MC) is $[0,1]$ whereas Pearson's ($r$) and Spearman's ($\rho$) coefficients range is $[-1,1]$.}
\label{tab:correlation_results}
\resizebox{\textwidth}{!}{%
\begin{tabular}{cccccccc}
\hline
\multicolumn{1}{|c|}{\multirow{2}{*}{\textbf{Category}}} & \multicolumn{1}{c|}{\multirow{2}{*}{\textbf{Feature}}} & \multicolumn{3}{c|}{\textbf{\begin{tabular}[c]{@{}c@{}}Standard Engagement \\ (Ad Clicks $<2,188, N = 2,854$)\end{tabular}}} & \multicolumn{3}{c|}{\textbf{\begin{tabular}[c]{@{}c@{}}High Engagement \\ (Ad Clicks $\geq 2,188, N = 432$)\end{tabular}}} \\
\multicolumn{1}{|c|}{} & \multicolumn{1}{c|}{} & \textbf{$r$} & \textbf{$\rho$} & \multicolumn{1}{c|}{\textbf{MC}} & \textbf{$r$} & \textbf{$\rho$} & \multicolumn{1}{c|}{\textbf{MC}} \\ \hline
\multicolumn{1}{|c|}{\multirow{3}{*}{\textit{Ad Metadata}}} & \multicolumn{1}{c|}{\textbf{Ad Impressions}} & 0.49*** & 0.94*** & \multicolumn{1}{c|}{0.93***} & 0.89*** & 0.76*** & \multicolumn{1}{c|}{0.89***} \\
\multicolumn{1}{|c|}{} & \multicolumn{1}{c|}{\textbf{Ad Lifetime}} & -0.05*** & 0.21*** & \multicolumn{1}{c|}{0.34***} & -0.03*** & -0.01*** & \multicolumn{1}{c|}{0.25***} \\
\multicolumn{1}{|c|}{} & \multicolumn{1}{c|}{\textbf{Ad Spend}} & 0.27*** & 0.79*** & \multicolumn{1}{c|}{0.70***} & 0.65*** & 0.23*** & \multicolumn{1}{c|}{0.70***} \\ \hline
\multicolumn{1}{|c|}{\multirow{2}{*}{\textit{Size}}} & \multicolumn{1}{c|}{\textbf{Character Count}} & -0.01*** & 0.08*** & \multicolumn{1}{c|}{0.22***} & n.s. & n.s. & \multicolumn{1}{c|}{0.21***} \\
\multicolumn{1}{|c|}{} & \multicolumn{1}{c|}{\textbf{Word Count}} & -0.02*** & 0.08*** & \multicolumn{1}{c|}{0.22***} & n.s. & n.s. & \multicolumn{1}{c|}{0.20***} \\ \hline
\multicolumn{1}{|c|}{\multirow{9}{*}{\textit{Sentiment \& Subjectivity}}} & \multicolumn{1}{c|}{\textbf{NLTK VADER Compound Score}} & n.s. & n.s. & \multicolumn{1}{c|}{0.15***} & n.s. & n.s. & \multicolumn{1}{c|}{0.17***} \\
\multicolumn{1}{|c|}{} & \multicolumn{1}{c|}{NLTK Negative Sentiment Only} & n.s. & n.s. & \multicolumn{1}{c|}{0.07***} & n.s. & n.s. & \multicolumn{1}{c|}{0.10*} \\
\multicolumn{1}{|c|}{} & \multicolumn{1}{c|}{NLTK Positive Sentiment Only} & n.s. & n.s. & \multicolumn{1}{c|}{0.04*} & n.s. & n.s. & \multicolumn{1}{c|}{0.17***} \\
\multicolumn{1}{|c|}{} & \multicolumn{1}{c|}{\textbf{Textblob Sentiment Polarity}} & n.s. & n.s. & \multicolumn{1}{c|}{0.08***} & n.s. & n.s. & \multicolumn{1}{c|}{0.15**} \\
\multicolumn{1}{|c|}{} & \multicolumn{1}{c|}{Textblob Negative Sentiment Only} & 0.01* & n.s. & \multicolumn{1}{c|}{0.07***} & n.s. & n.s. & \multicolumn{1}{c|}{n.s.} \\
\multicolumn{1}{|c|}{} & \multicolumn{1}{c|}{Textblob Positive Sentiment Only} & n.s. & n.s. & \multicolumn{1}{c|}{0.09***} & n.s. & n.s. & \multicolumn{1}{c|}{0.13**} \\
\multicolumn{1}{|c|}{} & \multicolumn{1}{c|}{\textbf{Textblob Subjectivity}} & 0.01* & n.s. & \multicolumn{1}{c|}{0.10***} & n.s. & n.s. & \multicolumn{1}{c|}{0.13**} \\
\multicolumn{1}{|c|}{} & \multicolumn{1}{c|}{Textblob Subjective Scores Only} & n.s. & n.s. & \multicolumn{1}{c|}{0.07***} & n.s. & n.s. & \multicolumn{1}{c|}{0.12*} \\
\multicolumn{1}{|c|}{} & \multicolumn{1}{c|}{Textblob Objective Scores Only} & n.s. & n.s. & \multicolumn{1}{c|}{0.05*} & n.s. & n.s. & \multicolumn{1}{c|}{0.11*} \\ \hline
\multicolumn{1}{|c|}{\multirow{4}{*}{\textit{\begin{tabular}[c]{@{}c@{}}LIWC Summary\\ Variables\end{tabular}}}} & \multicolumn{1}{c|}{\textbf{Authentic}} & n.s. & n.s. & \multicolumn{1}{c|}{0.08***} & n.s. & n.s. & \multicolumn{1}{c|}{0.15**} \\
\multicolumn{1}{|c|}{} & \multicolumn{1}{c|}{\textbf{Analytical Thinking}} & -0.09*** & -0.08*** & \multicolumn{1}{c|}{0.09***} & n.s. & n.s. & \multicolumn{1}{c|}{0.12*} \\
\multicolumn{1}{|c|}{} & \multicolumn{1}{c|}{\textbf{Clout}} & -0.06*** & -0.06** & \multicolumn{1}{c|}{0.08***} & n.s. & n.s. & \multicolumn{1}{c|}{0.16**} \\
\multicolumn{1}{|c|}{} & \multicolumn{1}{c|}{\textbf{Emotional Tone}} & n.s. & -0.04* & \multicolumn{1}{c|}{0.08***} & n.s. & n.s. & \multicolumn{1}{c|}{0.14**} \\ \hline
\multicolumn{1}{|c|}{\multirow{17}{*}{\textit{LIWC categories}}} & \multicolumn{1}{c|}{\textbf{Affective Processes}} & n.s. & n.s. & \multicolumn{1}{c|}{0.07***} & n.s. & n.s. & \multicolumn{1}{c|}{0.14**} \\
\multicolumn{1}{|c|}{} & \multicolumn{1}{c|}{\textbf{Social Processes}} & -0.03* & -0.04* & \multicolumn{1}{c|}{0.06**} & n.s. & n.s. & \multicolumn{1}{c|}{0.16**} \\
\multicolumn{1}{|c|}{} & \multicolumn{1}{c|}{\textbf{Cognitive Processes}} & 0.07*** & 0.06** & \multicolumn{1}{c|}{0.11***} & n.s. & n.s. & \multicolumn{1}{c|}{0.11*} \\
\multicolumn{1}{|c|}{} & \multicolumn{1}{c|}{\textbf{Perceptual Processes}} & n.s. & n.s. & \multicolumn{1}{c|}{0.17***} & n.s. & n.s. & \multicolumn{1}{c|}{n.s.} \\
\multicolumn{1}{|c|}{} & \multicolumn{1}{c|}{\textbf{Biological Processes}} & n.s. & n.s. & \multicolumn{1}{c|}{0.06**} & n.s. & n.s. & \multicolumn{1}{c|}{n.s.} \\
\multicolumn{1}{|c|}{} & \multicolumn{1}{c|}{\textbf{Drives}} & -0.10*** & -0.13*** & \multicolumn{1}{c|}{0.14***} & 0.15*** & 0.15** & \multicolumn{1}{c|}{0.21***} \\
\multicolumn{1}{|c|}{} & \multicolumn{1}{c|}{\textbf{Future Focus}} & 0.01*** & 0.08*** & \multicolumn{1}{c|}{0.11***} & n.s. & n.s. & \multicolumn{1}{c|}{n.s.} \\
\multicolumn{1}{|c|}{} & \multicolumn{1}{c|}{\textbf{Past Focus}} & 0.08*** & 0.13*** & \multicolumn{1}{c|}{0.14***} & n.s. & n.s. & \multicolumn{1}{c|}{0.14**} \\
\multicolumn{1}{|c|}{} & \multicolumn{1}{c|}{\textbf{Present Focus}} & n.s. & n.s. & \multicolumn{1}{c|}{0.11***} & n.s. & n.s. & \multicolumn{1}{c|}{0.15**} \\
\multicolumn{1}{|c|}{} & \multicolumn{1}{c|}{\textbf{Relativity}} & n.s. & n.s. & \multicolumn{1}{c|}{0.10***} & n.s. & n.s. & \multicolumn{1}{c|}{0.12*} \\
\multicolumn{1}{|c|}{} & \multicolumn{1}{c|}{\textbf{Work}} & 0.01*** & 0.10*** & \multicolumn{1}{c|}{0.18***} & n.s. & n.s. & \multicolumn{1}{c|}{0.11*} \\
\multicolumn{1}{|c|}{} & \multicolumn{1}{c|}{\textbf{Death}} & 0.03*** & 0.07*** & \multicolumn{1}{c|}{0.11***} & -0.04* & n.s. & \multicolumn{1}{c|}{n.s.} \\
\multicolumn{1}{|c|}{} & \multicolumn{1}{c|}{\textbf{Home}} & n.s. & n.s. & \multicolumn{1}{c|}{0.06**} & n.s. & n.s. & \multicolumn{1}{c|}{n.s.} \\
\multicolumn{1}{|c|}{} & \multicolumn{1}{c|}{\textbf{Leisure}} & -0.06* & -0.05** & \multicolumn{1}{c|}{0.07***} & n.s. & n.s. & \multicolumn{1}{c|}{n.s.} \\
\multicolumn{1}{|c|}{} & \multicolumn{1}{c|}{\textbf{Money}} & -0.09*** & -0.07*** & \multicolumn{1}{c|}{0.10***} & -0.08* & -0.10* & \multicolumn{1}{c|}{0.13**} \\
\multicolumn{1}{|c|}{} & \multicolumn{1}{c|}{\textbf{Religion}} & n.s. & n.s. & \multicolumn{1}{c|}{0.07***} & n.s. & n.s. & \multicolumn{1}{c|}{0.11*} \\
\multicolumn{1}{|c|}{} & \multicolumn{1}{c|}{\textbf{All Punctuation}} & n.s. & n.s. & \multicolumn{1}{c|}{0.10***} & n.s. & n.s. & \multicolumn{1}{c|}{n.s.} \\ \hline
\multicolumn{8}{l}{\begin{tabular}[c]{@{}l@{}}* Significant at p \textless .05\\ ** Significant at p \textless .01\\ *** Significant at  p \textless .001\\n.s. = not statistically significant\end{tabular}}
\end{tabular}%
}
\end{table}

As displayed in Table~\ref{tab:correlation_results}, we found moderate to strong positive correlations between Ad Clicks and the ad metadata features. 
For example, for both Standard and High Engagement groups, Ad Clicks strongly correlated with both amount of views an ad received ($\rho_{stand}=0.94,\ r_{high}=0.89,\ p < .001$) and ad expenditure ($\rho_{stand}=0.79,\ r_{high}=0.65,\ p < .001$). 
We also noticed that Ad Impressions was extremely similar in distribution to Ad Clicks, which is intuitive as views and clicks are both metrics of social media engagement~\cite{aldous2019view}. We therefore opted to discard the Ad Impressions feature from our analysis. 
Strong or moderate correlations did not hold true for the remaining feature categories. 
For example, we found nearly no statistically significant Pearson and Spearman correlations with sentiment and subjectivity features for both Standard and High Engagement groups. 
There were several trivial (i.e., $< 0.1$) correlations between Ad Clicks and sociolinguistic features for the Standard Engagement group, and no statistically significant results for the majority of the LIWC features for the High Engagement group.

\begin{figure}
    \includegraphics[width=\linewidth]{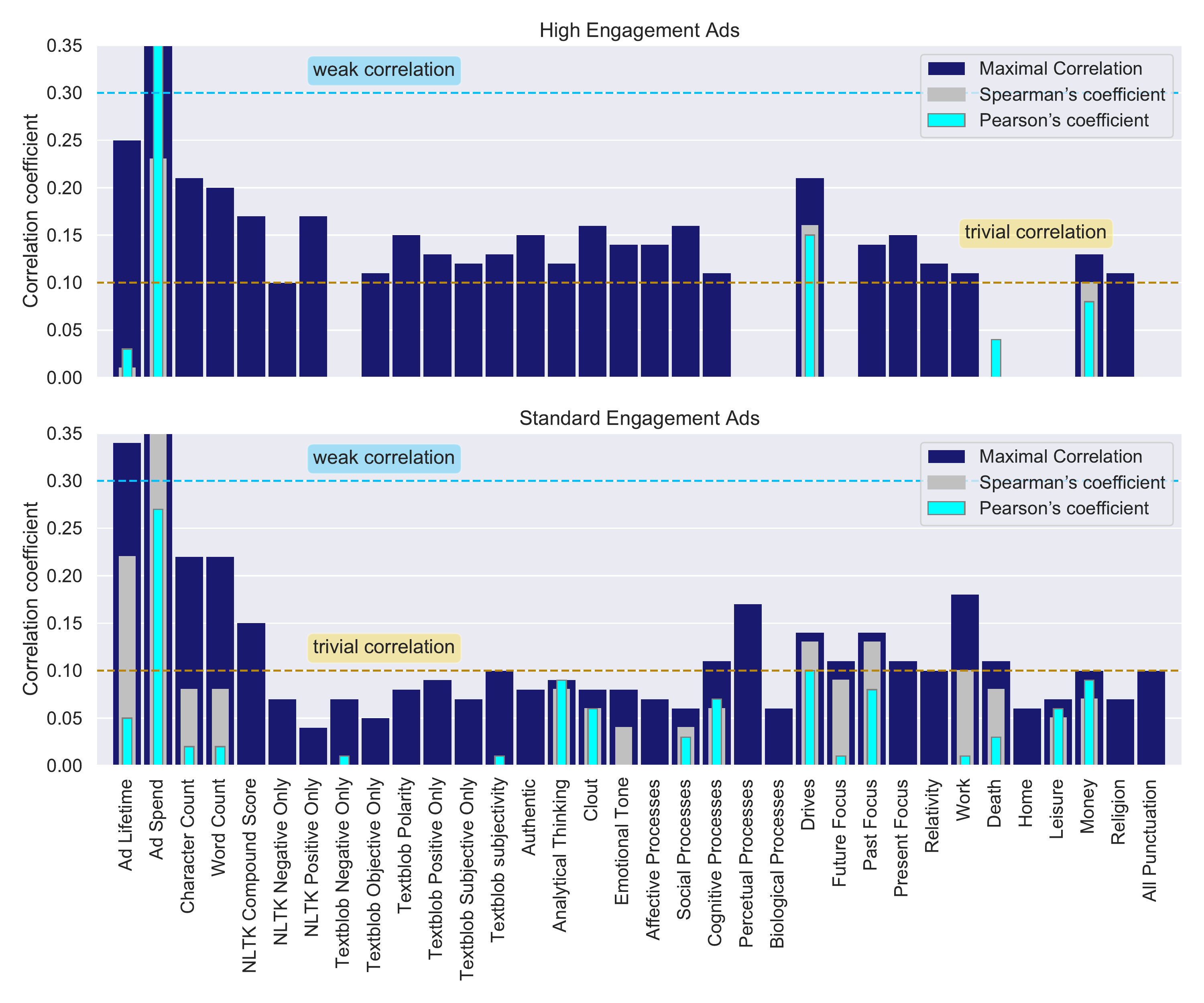}
    \caption{Correlation results (absolute values) for Ad Clicks vs. features for High Engagement ads (upper panel) and Standard Engagement ads (lower panel).}
    \label{fig:correlation_results}
\end{figure}

Note that, in terms of magnitude, the majority of the Spearman's coefficients were larger than Pearson's coefficients for the Standard Engagement group, indicating a more non-linear behavior, and most of the High Engagement group exhibited no statistically significant results. 
Nonetheless, these overall low correlation coefficients can be explained based on the number of trivial correlations ($\rho, r < 0.1$), indicating that these variables do not exhibit a monotonic nor a linear relationship, and therefore $\rho$ and $r$ cannot fully describe their pairwise correlations. 
We thus used the Alternating Conditional Expectations (ACE) algorithm to find the fixed point of Maximal Correlation (MC) for each feature; in other words, we transform the dependent and independent variables to maximize Pearson's correlation coefficient between the transformed dependent and transformed independent variables. 
Deebani and Kachouie et al.~\cite{Deebani2018-rt} tested several correlation analysis methods on several simulations of different relationship types, with and without noise, and found that Maximal Correlation equaled or outperformed Pearson's and Spearman's correlation.
The authors thus describe Maximal Correlation as efficient and robust to noise, and allows for non-linear correlations to be detected. 
It is important to note that MC ranges from $[0, 1]$, i.e., it does not measure the polarity of the correlation.
Nonetheless, as MC is able to capture both linear and non-linear relationships, this transformation results in greater predictive value in the extracted features, as displayed in Fig.~\ref{fig:correlation_results}. 

\subsubsection{Ad Metadata}
Maximal Correlation greatly improved the Pearson's correlation between the metadata features and Ad Clicks. 
For example, Ad Lifetime exhibited a trivial linear (i.e., Pearson) correlation with clicks for both Standard and High Engagement groups ($r_{stand}= -0.05, r_{high}=-0.03, p <.001$); these values increased to a weak to moderate relationship following the ACE transformation: $MC_{stand} = 0.34, MC_{high}=0.25, p<.001$. 
Similarly, Ad Spend's Maximal Correlation $MC_{stand}=0.70, p<.001$ for the Standard Engagement group better reflects the strong positive Spearman correlation ($\rho_{stand}=0.79, p<.001$) instead of the weak positive Pearson correlation ($r_{stand}=0.27, p<.001$).

\subsubsection{Text Size}
Prior to the Maximal Correlation transformations, the character and word counts of Standard Engagement ads exhibited near-zero linear correlation with Ad Clicks ($r_{stand}=-0.01, r_{stand}=-0.02, p<.001$, respectively); this increased to a small positive correlation ($MC_{stand}=0.22, p<.001$ for both character and word counts) following the Maximal Correlation transformations.
The results for the High Engagement group also greatly improved from not statistically significant for both character and word count, to $MC_{high} = 0.21$ and $MC_{high}=0.20$ ($p<.001$), respectively. 

\subsubsection{Sentiment \& Subjectivity}
Arguably one of the largest improvements was seen with NLTK and TextBlob sentiment and subjectivity scores based on Maximal Correlation.
High Engagement ads demonstrated small Maximal Correlations (ranging from $0.10$ to $0.17$, $p < .05$) for nearly every feature, whereas for Standard Engagement ads, only the NLTK Compound Score and TextBlob Subjectivity Score demonstrated small MC correlations ($0.15$ and $0.10, p < .001$, respectively). 
Like previously stated, Maximal Correlation does not report the direction of the relationship between the variables; 
however, NLTK Compound Score and NLTK Negative Sentiment Only score also exhibited the largest mean differences between the Standard and High Engagement groups: $\mu_{stand}=0.08$ vs. $\mu_{high}=0.17$ for NLTK Compound Score, and $\mu_{stand}=-0.24$ vs. $\mu_{high}=-0.12$, respectively. 
Therefore, not only did High Engagement ads demonstrate higher (in terms of polarity and magnitude) Maximal Correlations with sentiment and subjectivity than the Standard Engagement ads, but High Engagement ads were also more positive in sentiment, on average, than Standard Engagement ads.

Additionally, sentiment and subjectivity variables were further analyzed using the Chi-Squared test (see Table~\ref{tab:chisq_results}). 
Features were transformed into categorical variables using the thresholds described in Sec.~\ref{sec:methodology_sentiment} (e.g., NLTK Compound scores $\geq 0.05$ were labeled as positive). 
All sentiment features (VADER, TextBlob, and Flair) were found to be dependent with engagement as measured by the Standard and High Engagement Ad Click groups ($p<.001$), suggesting that sentiment features are associated with ad engagement. 
However, subjectivity was not statistically significant and therefore TextBlob Subjectivity was independent of engagement ($\chi^2 (1, N=3,286) = 1.574, p = 0.21$).  
Moreover, Maximal Correlation was statistically significant for both groups (High vs. Standard Engagement), further emphasizing the robustness of this correlation analysis method in contrast to Pearson's and Spearman's correlations. 

\begin{table}[h]
\caption{Chi-Squared tests analyzing Ad Clicks (for Standard and High Engagement) vs. sentiment and subjectivity features.}
\label{tab:chisq_results}
\resizebox{0.70\textwidth}{!}{%
\begin{tabular}{cccccc}
\toprule
\textbf{Feature Category} & \textbf{Feature} & \textbf{$\chi^2$ Stat} & \textbf{df} & \textbf{N} & \textbf{p-value} \\ 
\midrule
\multirow{4}{*}{\textit{Sentiment \& Subjectivity}} & \textbf{NLTK VADER Compound Score} & 54.122 & 2 & 3,286 & \textless .001 \\ \cline{2-6} 
 & \textbf{TextBlob Sentiment Polarity} & 31.749 & 2 & 3,286 & \textless .001 \\ \cline{2-6} 
 & \textbf{Flair Sentiment} & 17.965 & 1 & 3,286 & \textless .001 \\ \cline{2-6} 
 & \textbf{TextBlob Subjectivity} & 1.574 & 1 & 3,286 & 0.210 \\
 \bottomrule
\end{tabular}%
}
\end{table}

\subsubsection{Sociolinguistic Features}\label{sec:correlation_liwc}
For the four summary LIWC variables, the High Engagement group showed weak Maximal Correlations (ranging from $0.12$ to $0.16, p<.05$) whereas Standard Engagement showed only trivial ($< 0.1, p<.001$) correlations. 
Interestingly, the mean values for the summary variables were nearly unchanged across the engagement groups, with Analytic as the exception: the average score drops from $70.27$ to $60.10$ for the Standard vs. High Engagement groups suggesting that High Engagement ads were more informal and personal than Standard Engagement ads.

The remaining LIWC categories experienced nearly no variation in terms of mean values across the Standard and High Engagement groups. 
Six features (Perceptual Processes, Biological Processes, Future Focus, Death Home, Leisure, and All Punctuation) were not found to be statistically significant regardless of correlation analysis method for High Engagement ads, whereas all LIWC features were found to be statistically significant ($p<.01$) for the Standard Engagement group (though six features exhibited trivial MC coefficients). 
This could be due to the discrepancy between the average text length for Standard and High Engagement ads: 45 words vs. 26 words---it is possible that more statistically significant results could not be found for the High Engagement group due to low sample size.

Additionally, the feature Drives (a LIWC dimension that captures motive and needs, such as risk and rewards) stands out, as it was the only LIWC feature to find non-trivial (i.e., $\geq 0.1$) Pearson, Spearman, and Maximal Correlations for both Standard and High Engagement. 
Both engagement groups experienced nearly the same average Drives value ($\mu \approx 14$, i.e., $\sim 14\%$ of ad texts contained Drives-related works) and this average value was the third largest mean LIWC category value for the High Engagement group. 
Since we did not capture the sub-categories of the Drives dimension (i.e., Affiliation, Achievement, Power, Reward Focus, and Risk Focus), we thus can only conclude that Drives appears to be associated with ad engagement (corroborated by the results in Sec.~\ref{sec:results_feature_selection}), especially for High Engagement ads.
\\

\takeaway{(\rqone) Ad expenditure followed by ad lifetime and text size showed the highest Maximal Correlations for both Standard and High Engagement groups. 
High Engagement ads were more positive in terms of sentiment, more informal and personal, and shorter in size than Standard Engagement ads.}

\subsection{Feature Selection (\rqtwo\ \& \rqthree: Features Predicting Ad Engagement)}\label{sec:results_feature_selection}


Our correlation and statistical analyses used to address \rqone\ relied on the individual relevance of each feature in characterizing engagement. However, individual features sometimes fail in predicting the target variable accurately. Machine learning models can combine multiple features to predict the target, sometimes revealing promising features that do not have relevant pairwise correlation results. In this section, we present and discuss our investigative steps and results to address \rqtwo\ and \rqthree, where we aimed to 
determine the set of features (from Table~\ref{tab:dataset_info}) that best predict engagement via machine learning analysis. 
There are many feature selection techniques available in the literature~\cite{Theodoridis2008-me}. 
In this work, we use Recursive Feature Elimination (RFE) to determine from our set of collected features (Table~\ref{tab:dataset_info}) which subset should be retained to best predict ad engagement. 
RFE employs a multi-class classifier as estimator to rank the relevance of existing features by assigning a weight coefficient to determine their importance and select the optimal feature subset (i.e., the subset with best prediction results) in a supervised fashion~\cite{Theodoridis2008-me}. 
As RFE weighs the features based on the importance in predicting engagement, it can become sensitive to the type of model used.
For this reason, we performed this analysis for six different estimators and then compared their results, checking for commonalities among the selected feature subsets. Six popular classifiers were used, in particular: Adaboost, Bernoulli Naive Bayes (NB), Gradient Boosting, Support Vector Machine with a linear kernel (Linear SVM), Logistic Regression, and Random Forest (see Tables~\ref{tab:feature_importance_summary} and \ref{tab:top_5_feature_selection}). 
We used their implementations available in the scikit-learn library~\cite{pedregosa2011scikit} (a popular machine learning library for Python) with default parameters in all cases. 
RFE was performed using stratified 5-fold cross-validation due to its relatively low bias and variance~\cite{han2011data}.
The evaluation metric used to rank the features and select the optimal feature subset was the F-score (the harmonic mean of precision, a measure of \textit{exactness}, and recall, a measure of \textit{completeness}~\cite{han2011data}), which is well suited to handle imbalanced datasets as in our case ($2,854$ Standard vs. $432$ High Engagement ads)~\cite{han2011data}. Importantly, prior to this analysis, all features were first standardized by removing the mean and scaling to unit variance. Our results are summarized in 
in Table~\ref{tab:feature_importance_summary}, for each classifier, we present the average F-score computed among all subsets of features tested (containing $n=\{k,k-1,k-2,...1\}$ features where $k=\totalfeat$, the total number of features), as well as the lowest and highest F-score achieved, including the size of the optimal feature subset associated with the highest F-score.

\begin{table}[h]
\caption{Summary of the performance of all models used for feature selection, ranked based on mean F-score.}
\label{tab:feature_importance_summary}
\resizebox{0.7\textwidth}{!}{%
\begin{tabular}{ccccccc}
\toprule
\multirow{2}{*}{\textbf{Rank}} & \multirow{2}{*}{\textbf{Classifier}} & \multicolumn{4}{c}{\textbf{F-Score}} & \multirow{2}{*}{\textbf{Optimal \# of Features}} \\
 &  & \textbf{Mean} & \textbf{$\sigma$} & \textbf{Min} & \textbf{Max} &  \\ \midrule
1 & Linear SVM & 93.6\% & 0.0\% & 93.6\% & 93.7\% & 12 \\
2 & Logistic Regression & 93.6\% & 0.1\% & 93.4\% & 93.8\% & 6 \\
3 & Gradient Boosting & 93.4\% & 0.1\% & 93.1\% & 93.8\% & 3 \\
4 & Random Forest & 93.0\% & 0.1\% & 93.0\% & 93.7\% & 1 \\
5 & Adaboost & 93.0\% & 0.2\% & 92.7\% & 93.6\% & 4 \\
6 & Bernoulli NB & 90.0\% & 2.8\% & 84.1\% & 93.0\% & 1 \\
\bottomrule
\end{tabular}%
}
\end{table}

\subsubsection{Ad Metadata}
The metadata features mirrored the results from the Maximal Correlation analysis (\rqone), that is, both Ad Spend and Lifetime were selected as important features for predicting engagement, particularly in distinguishing between standard and high engagement.
Ad expenditure was ranked the most important in all but one of the six models used, with Bernoulli NB nonetheless ranking it as the second most important feature. 
Similarly, the lifetime of the ad was ranked top 5 by all except the Logistic Regression model. 

\subsubsection{Text Size}
Similarly, the size of the ad's text as measured by word or character count appeared in the top 2 most important features for all models tested. 
This again corroborated our Maximal Correlation results, wherein text size features exhibited the second largest Maximal Correlation values for both Standard and High Engagement ads. 
Based on the average values for text size, we can infer that shorter ad texts were more engaging (e.g., character count: $\mu_{stand}=270.20, \mu_{high}=163$). 

\subsubsection{Sentiment \& Subjectivity}
At least one sentiment or subjectivity feature was ranked top 5 by five of the six models. 
Notably, NLTK's VADER Compound Score was ranked first by three models: Linear SVM and Adaboost, and Linear SVM, whereas NLTK Positive Only scores and NLTK Negative Only scores were selected by both Linear SVM and Logistic Regression.
Based on Table~\ref{tab:dataset_info}, we see that NLTK Negative scores were more negative for Standard Engagement than for High Engagement ($\mu_{stand}=-0.24, \mu_{high}=-.12$), and NLTK Compound Scores were more positive more High Engagement ads ($\mu_{stand}=0.08, \mu_{high}=0.17$), which is in accordance with our earlier observations that positive sentiment disinformation ads in this dataset were more engaging.

\subsubsection{Sociolinguistic Features}
Two summary variables appeared in three out of the six models tested for feature selection: Analytic (Linear SVM, Logistic Regression, and Gradient Boosting) and Authentic (Linear SVM and Logistic Regression).
This elucidates and stands with our results in Sec.~\ref{sec:correlation_liwc}: High Engagement ads were more informal and personal than Standard Engagement ads.

Two other LIWC features were selected among the top 5 by the feature selection models: Drives (captures motives, drives, and needs, \eg risk, rewards) and Religion (captures religion-related words), chosen by 3 and 4 of the 6 models, respectively.
Whereas Drives experienced nearly the same min, mean, and max values for both High and Standard Engagement, Religion differed in max values for Standard ($max = 66$) and High ($max = 33$); therefore, Religion showed greater variety and range of values for the Standard group, yet all Religion scores for both Standard and High Engagement group were relatively low ($\mu_{stand}=0.98, \mu_{high}=0.64$), indicating low use of religious language across all ads.
Conversely, both Drives ($MC_{stand}=0.14, MC_{high}=0.21, p<.001$) and Religion ($MC_{stand}=0.07, p<.001; MC_{high}=0.11, p<.05$) showed higher Maximal Correlation values for the High Engagement group as compared to Standard Engagement. 
Most notably, LIWC features largely dominated four of the six feature selection models: 8/16 for Linear SVM, 5/10 for Logistic Regression, 4/8 for Adaboost, and 3/5 for Bernoulli NB. 
\\

\takeaway{(\rqtwo) Ad expenditure was ranked as the \#1 most important feature by all 6 models, followed by text size, ad lifetime, and NLTK's VADER Compound Score (sentiment measure) appearing as recurring important features across the models. LIWC's sociolinguistic features made up at least half of the top 5 most important features for four of the six feature selection models (Linear SVM, Logistic Regression, Adaboost, and Bernoulli NB).}

\takeaway{
(\rqthree) The Linear SVM and Logistic Regression classifiers exhibited the highest average F-scores (93.6\% in both cases) in accurately classifying an ad as High or Standard Engagement. SVM includes a wider range of explainable features, allowing for a better determination of engagement in a disinformation ad.
}


\begin{table}[h]
\caption{Top 5 ranked features for each model tested for feature selection.}
\label{tab:top_5_feature_selection}
\resizebox{\textwidth}{!}{%
\begin{tabular}{lcccccc}
\toprule
\textbf{Rank} & \textbf{Linear SVM} & \textbf{Logistic Regression} & \textbf{Gradient Boosting} & \textbf{Random Forest} & \textbf{Adaboost} & \textbf{Bernoulli NB} \\ \midrule
1 & \begin{tabular}[c]{@{}c@{}}Ad Spend\\ Religion\\ Drives\\ Biological Processes\\ NLTK Negative Only\\ Authentic\\ Analytical Thinking\\ Emotional Tone\\ NLTK Compound\\ NLTK Positive Only\\ Character Count\\ Word Count\end{tabular} & \begin{tabular}[c]{@{}c@{}}Ad Spend\\ Word Count\\ NLTK Negative Only\\ Analytical Thinking\\ Character Count\\ NLTK Positive Only\end{tabular} & \begin{tabular}[c]{@{}c@{}}Ad Spend\\ Character Count\\ Ad Lifetime\end{tabular} & Ad Spend & \begin{tabular}[c]{@{}c@{}}Ad Spend\\ Character Count\\ NLTK Compound\\ Ad Lifetime\end{tabular} & Religion \\ \hline
2 & Past Focus & Religion & Word Count & Character Count & All Punctuation & Ad Spend \\ \midrule
3 & Affective Processing & Drives & Analytical Thinking & Word Count & Religion & Home \\ \midrule
4 & Ad Lifetime & Authentic & NLTK Compound & Drives & Cognitive Processes & Ad Lifetime \\ \midrule
5 & TextBlob Objective Only & Biological Processes & All Punctuation & Ad Lifetime & Past Focus & Death \\ \bottomrule
\end{tabular}%
}
\end{table}

\subsection{Topic Modeling}\label{sec:topic_modeling}
Topic modeling was performed on the ads' text using Latent Dirichlet Allocation (LDA), an unsupervised probabilistic generative model.
Simple textual preprocessing was done to make the text more amenable for analyses, including the removal of punctuation and stop words, and the lowercasing of all words.
In order to transform the textual data into a format that serves as input for the LDA model, we converted the texts into a simple vector representation using bag of words (BoW).
Then, we converted the list of ad texts into lists of vectors, all with length equal to the vocabulary. 
Words were then lemmatized, keeping only nouns, adjectives, verbs, and adverbs.

We validated the LDA's topic modeling performance using topic coherence, as described in \cite{Roder2015-kj}, and is made readily available in \texttt{gensim.models} module for Python.
Using the $c_v$ coherence, i.e., the coherence computed as the average similarity between the top word context vectors and their centroid, we find the set of parameters with maximum coherence value of $0.58$ for the entire dataset: $\beta = 0.01$ and $\alpha = 0.91$, yielding a total of $8$ topics. 
Using these parameters to train the LDA model, we then reduced the number of repeated keywords across different topics---i.e., each topic should describe a unique idea.

The groups of people that were targeted for each advertisement was provided amongst the several metadata provided in our original dataset. 
Using this information and the keywords associated with each topic, we then inspected the cleaned LDA topic results and proposed topic labels; for example, keywords such as \textit{conservatism, republican, tea party, confederate, Fox News, Trump, Pence, conservative} were assigned to the ``conservative or Republican'' category.
Therefore, we proposed the following eight overarching topic categories: 
(1) American patriotism, 
(2) justice/African-American, 
(3) perseverance/liberal/democrat, 
(4) female rights/education, 
(5) peace/guns, 
(6) police/military, 
(7) community integration/LGBT, and
(8) capitalism/conservative/republican. 
A summary of these results, along with example keywords, can be found in Table~\ref{tab:topic modeling}.

\begin{table}[h]
\caption{Example keywords with the largest weight contribution to each topic. Topics were based on LDA topic modeling.}
\label{tab:topic modeling}
\resizebox{\textwidth}{!}{%
\begin{tabular}{rccl}
\toprule
\multirow{2}{*}{\textbf{Proposed Summary Topic}} & \multicolumn{2}{c}{\textbf{N}} & \multirow{2}{*}{\textbf{Example keywords}} \\
 & \textbf{Standard Engagement} & \textbf{High Engagement} &  \\ \midrule
\textbf{(1)} American patriotism & 411 & 70 & support, follow, vote, go, veteran, always, give \\
\textbf{(2)} Justice and African American rights & 333 & 53 & justice, year, group, racism \\
\textbf{(3)} Perseverance, liberal political movement, Democratic Party & 376 & 51 & fight, take \\
\textbf{(4)} Female rights and education & 261 & 34 & woman, student, arrest \\
\textbf{(5)} Peace and guns & 308 & 54 & let, think, war, need, right \\
\textbf{(6)} Police/military & 360 & 81 & cop, life, brutality, shoot, video \\
\textbf{(7)} Community integration, LGBT rights & 286 & 63 & stand, stay, nation, proud \\
\textbf{(8)} Capitalism, conservative political movement, Republican Party & 490 & 55 & free, self-defense, class, safe, world, white \\ \midrule
\textbf{Total} & 2,854 & 432 & \\ \bottomrule
\end{tabular}%
}
\end{table}

\begin{figure}
    \includegraphics[width=\linewidth]{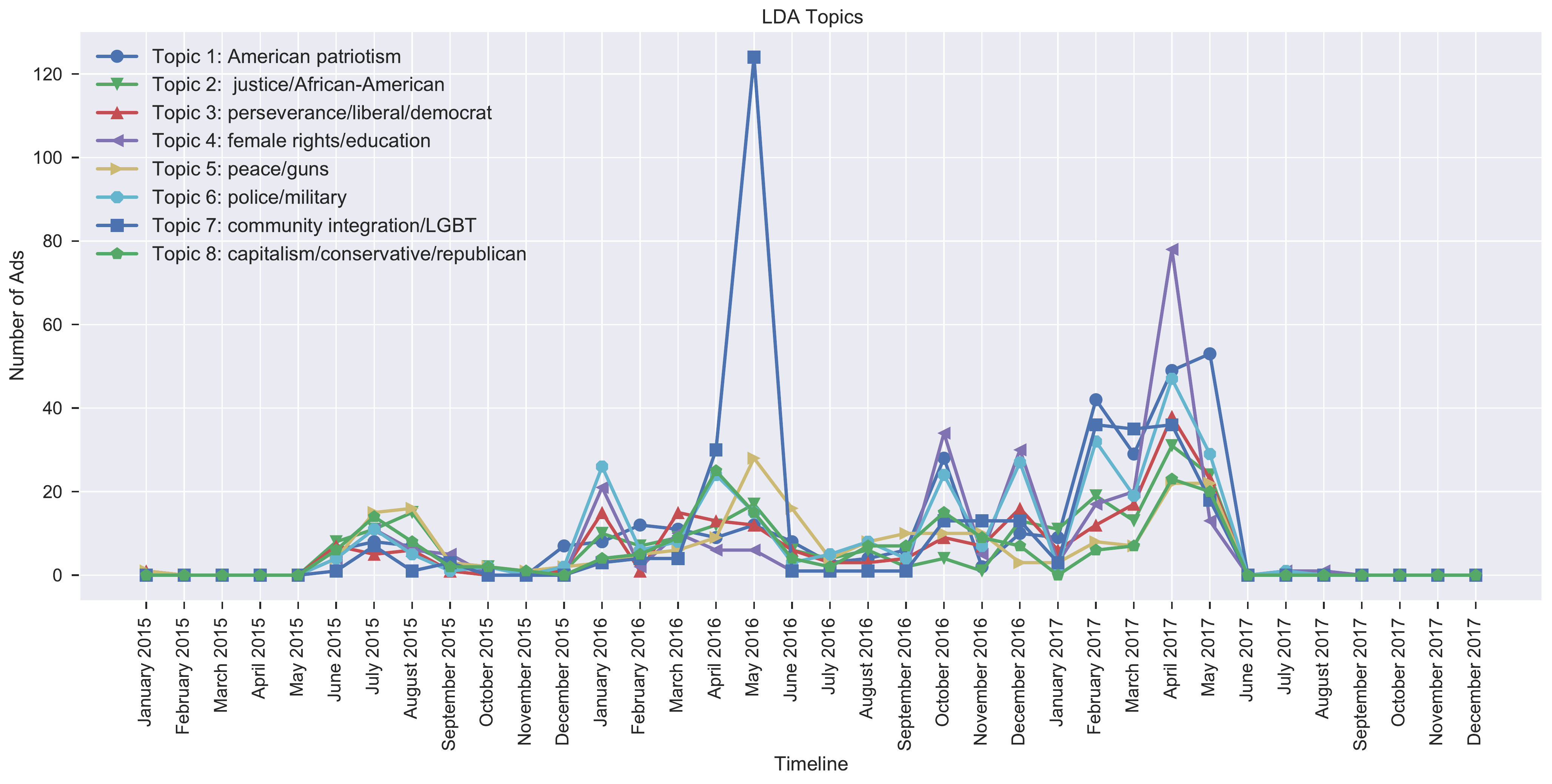}
    \caption{Total number of ads for each summary topic predicted by LDA.}
    \label{fig:LDATopics_TotalAdClicks_Timeline}
\end{figure}

\begin{figure}
    \includegraphics[width=\linewidth]{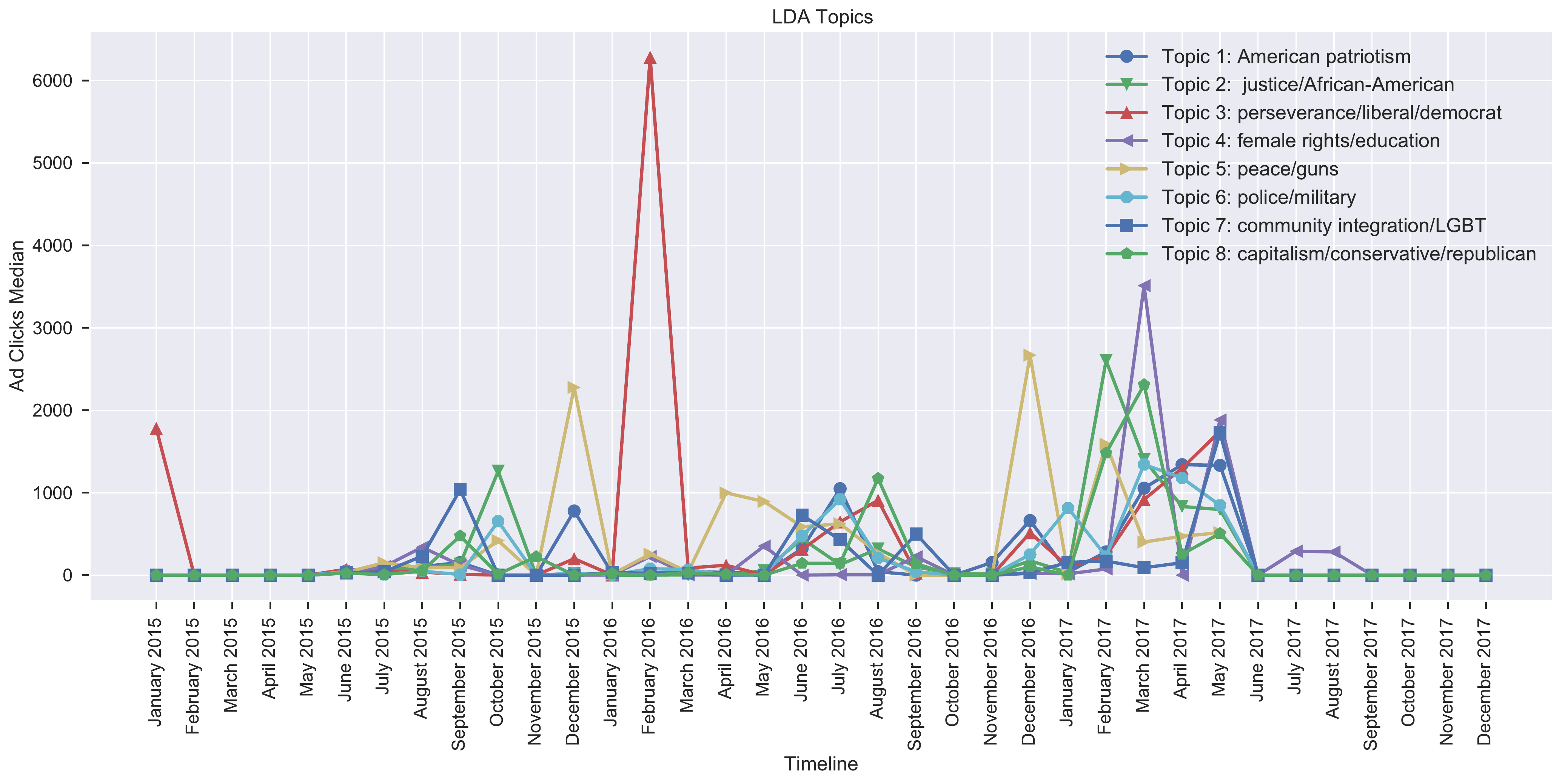}
    \caption{Median Ad Clicks for each summary topic predicted by LDA.}
    \label{fig:LDATopics_MedianAdClicks_Timeline}
\end{figure}

Our results, as analyzed and validated using Machine Learning algorithms, are in agreement with the qualitative analyses presented in prior works~\cite{DiResta_IRA_report, howard18}.
Figure~\ref{fig:LDATopics_TotalAdClicks_Timeline} shows the occurrence of each summary topic derived by the LDA topic modeling from June 2015 to August 2017. 
We see that the majority of Topic 7 (community integration/LGBT) has the largest ad count preceding the election (May 2016).
Interestingly, Topic 3 (perseverance/liberal/democrat) closely mirrors Topic 8 (capitalism/conservative/republican). 
We also observe several interesting occurrences when considering the median number of ad clicks for each summary topic during this same time period (Fig.~\ref{fig:LDATopics_MedianAdClicks_Timeline}).
Topic 3 (perseverance/liberal/democrat) stands out in engagement before the election (February--July, 2016) as well as significant impact during office takeover.
Topic 5 (peace/guns) has some significant engagement in the months preceding the election and some impact during office takeover. 
Topic 1 (American patriotism) and Topic 7 (community integration/LGBT) appear to follow each other throughout this timeline.
Topic 2 (justice/African-American) experiences relatively low median engagement numbers with the exception of a spike during the office takeover period. 
In January 2017, there was a surprisingly big significant spike in engagement in both figures. 
\\

\takeaway{(\rqfour) We confirm prior works by DiResta et al.~\cite{DiResta_IRA_report} and Howard et al.~\cite{howard18}, wherein the IRA purposefully targeted racial, ethnic, and political communities within the U.S. to further polarize political discourse. During key moments of the 2016 presidential election (e.g., during President Trump's office takeover, circa February--May 2017), several communities (e.g., African Americans, LGBT) experienced a surge in engagement with the IRA disinformation ads in our dataset.}
\section{Discussion}\label{sec:discussion}
We sought out to investigate several research questions pertaining to engagement in a dataset of Facebook ads created by the IRA during Russia's latest active measures campaign perpetrated before and after the 2016 U.S. presidential election, with the goal to influence the election results and sow discord in American citizens over divisive societal issues. 
To do so, we leveraged descriptive statistical and machine learning analyses to explore a total of \totalfeat\ features extracted and computed from the dataset. 
Engagement was defined as clicks on the ad because other engagement metrics (e.g., likes, shares) were not available in the dataset curated by Facebook.
This section analyzes our findings and the limitations of our work.

\subsection{Predictors of Engagement}

\subsubsection{Ad Lifetime and Expenditure}
Our Maximal Correlation results (Table~\ref{tab:correlation_results}) show that ad lifetime and expenditure has a moderate to strong relationship with both Standard and High Engagement. 
This was supported during our feature selection analysis.
The average Ad Spend for the High Engagement group was notably higher than that of Standard Engagement group (7,311 vs. 917 RUB). 
We hypothesize an intuitive explanation: paying more for an ad might be associated with a better targeting service from social media platforms, potentially causing the ad to reach more people who will be more interested in the ad (and views are a direct precursor to Ad Clicks). 
We also hypothesize that engagement potentially occurs as soon as users view an ad, and extending the ad's lifetime will likely not alter how users perceive and interact with the ad.

\subsubsection{Text Size}
\begin{figure}
    \centering
    \begin{subfigure}[b]{0.3\textwidth}
            \includegraphics[width=0.8\linewidth, page=2, trim={4.5cm 8.5cm 4.5cm 1.5cm}, clip]{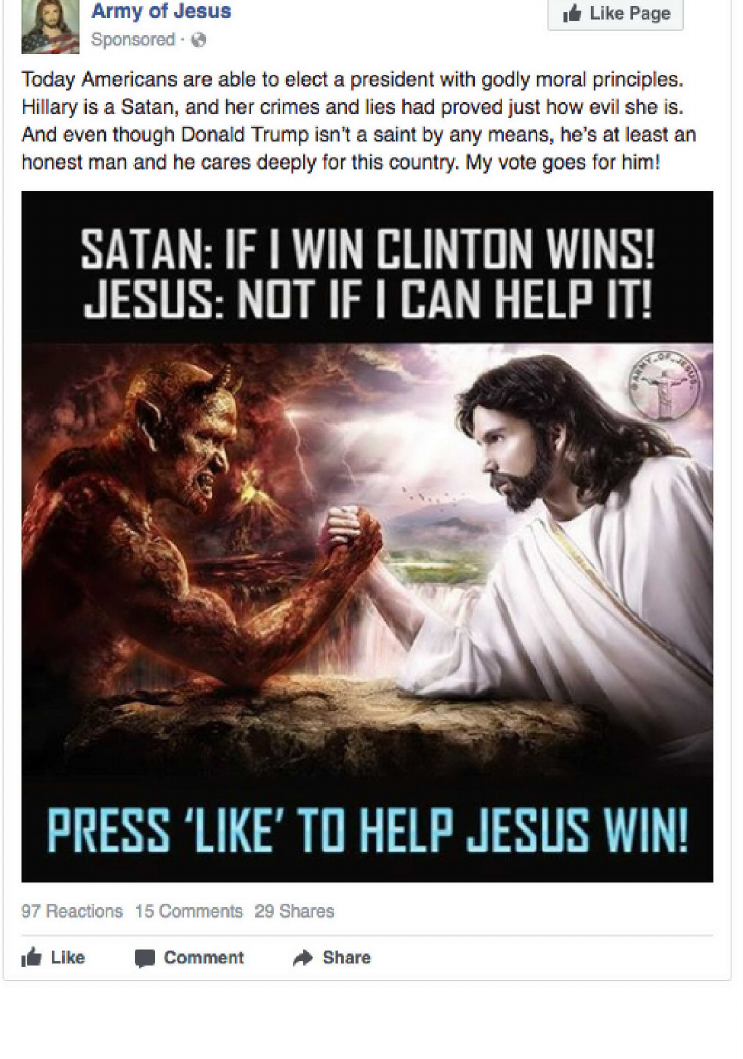}
    \end{subfigure}
    \begin{subfigure}[b]{0.3\textwidth}
            \includegraphics[width=0.8\linewidth, page=2, trim={4.5cm 8.5cm 4.5cm 1.5cm}, clip]{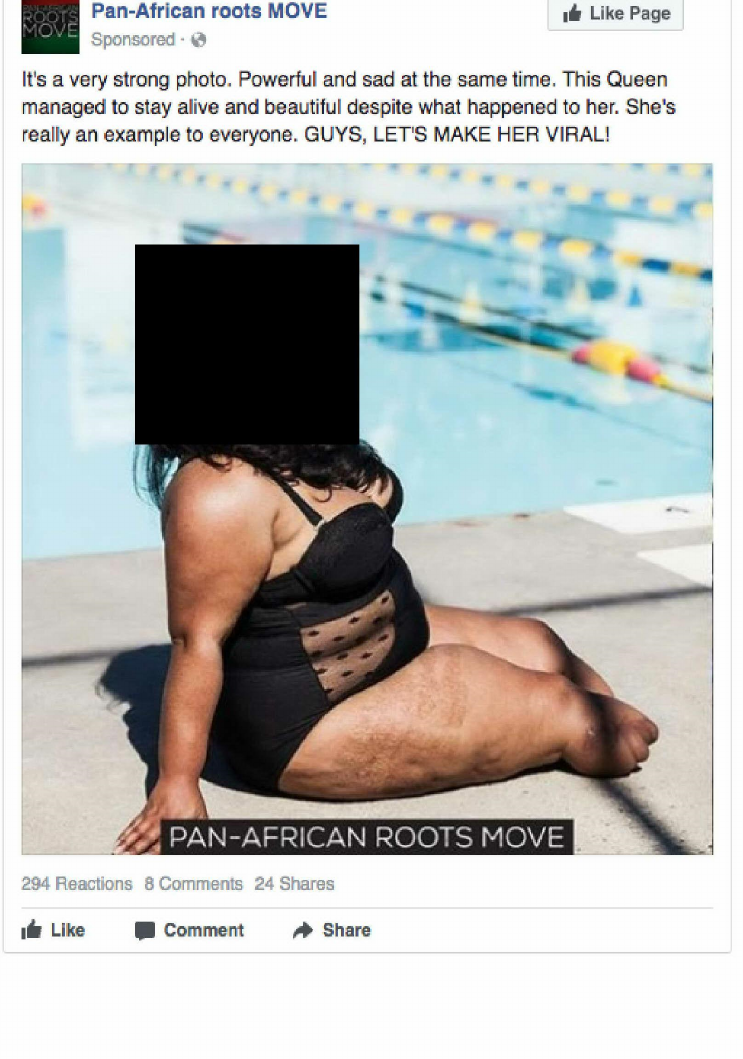}
    \end{subfigure}
    \begin{subfigure}[b]{0.3\textwidth}
        \includegraphics[width=0.8\linewidth, page=2, trim={4.5cm 8.5cm 4.5cm 1.5cm}, clip]{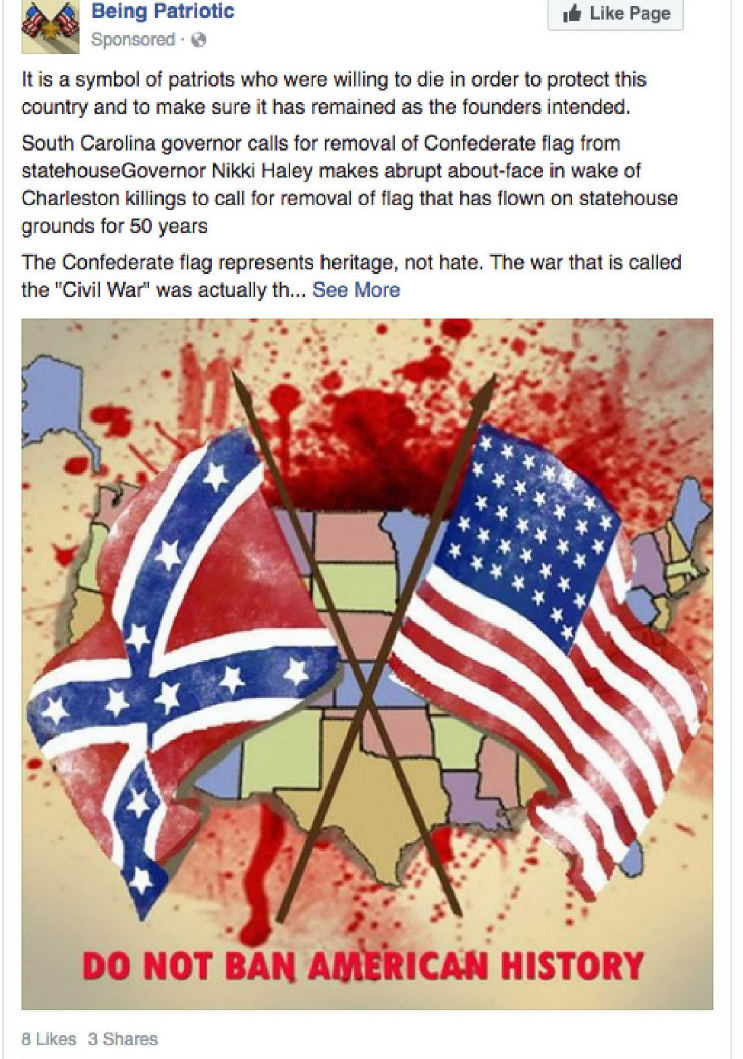}
    \end{subfigure}
    \caption{Examples of emotional or visceral images included in the IRA ads.}
    \label{fig:visceral_images}
\end{figure}

Another important feature was the length of the ad's text.
Facebook truncates posts greater than 477 characters~\cite{Gessler2016-nm}.
High Engagement ads had, on average, nearly 110 fewer characters (and nearly 20 fewer words) than the Standard Engagement group; therefore, we hypothesize that shorter ads are more engaging. 
Research on deception detection shows that deceivers embed influence cues in their content to blur people’s decision making~\cite{Kahneman2011-sq}.
In fact, accounts from Cold War disinformation points to the use of pictures, short texts, sexual appeal (if applicable), sensationalism, and high-arousal emotion in disinformation stimuli~\cite{Rid2020-bv}.
However, we only considered the textual content of each ad and disregarded the presence of images.
It is possible that ads with shorter texts used emotionally visceral images (examples in Fig.~\ref{fig:visceral_images}) to communicate a message, likely increasing users' engagement advertisement. 

\subsubsection{Sentiment \& Subjectivity}

We found that sentiment features were highly important for predicting engagement, with High Engagement ads more positive in sentiment than Standard Engagement ads.
Corroborating this finding, there is indeed a wealth of cognitive and behavioral sciences research that points to the impact of affective states (i.e., emotions) in decision making~\cite{Forgas2001-rg, Isen1991-yd}, where positive emotions have been shown to be more detrimental to rational decision-making than negative emotions.
Positive affect states have been shown to cause an increase in trust and a decrease in social vigilance~\cite{Kahneman2011-sq,Kircanski18}; therefore, a user's good mood indicates a safe environment~\cite{Kahneman2011-sq}, and can thus increase one's susceptibility to deception. 
Several works have also shown that high emotional arousal is leveraged by con artists to persuade victims to comply with their requests~\cite{Loewenstein1996-oz, Kircanski18} by focusing the victims' attention onto reward cues~\cite{Langenderfer2001-nv}.

\subsubsection{Sociolinguistic Features}
The LIWC sociolinguistic features are separated into two broad categories: summary variables (Analytic, Authentic, Clout, and Tone) and other LIWC categories (e.g., cognitive processes). 
For the summary variables, we found that High Engagement ads were more personal and informal than Standard Engagement ads.
Evans and Krueger~\cite{Evans2009-hp} and Cialdini's principles of persuasion~\cite{Cialdini2006-dk} offer plausible explanations to this: people who are perceived as familiar or similar (e.g., same culture) are more likely to be trusted by others (a phenomenon termed the \textit{in-group trust disposition}) and are more likely to have their requests obeyed. 
Therefore, ads whose authors masqueraded themselves as part of the targeted community may have achieved higher engagement levels.


In total, LIWC features largely dominated four of the six feature selection models.
From this, we see that the content of the advertisement itself, along with the use of (or lack thereof) certain topics (e.g., religion) influences engagement. 
Furthermore, prior works (e.g., \cite{silva2020predicting, howard18}) have shown that the type of user account (e.g., bot vs. real human Twitter accounts) impacts user engagement with a message. 
In this paper, we found that LIWC features such as Authentic (which measures how authentic a writer appears to be) suggest that the authors of the Facebook ads may have impacted users' engagement (e.g., an African American user posting about \#BlackLivesMatter).
The IRA has been shown to groom real users~\cite{Schifrin2020-ru} into writing their disinformation articles; as such, future works should analyze the accounts of users responsible for spreading the disinformation in our dataset.
Doing so may increase our awareness of the IRA's modus operandi.





\subsection{Limitations \& Future Works}\label{sec:limitations}
We now discuss this paper's limitations with an eye towards potential future works.

\subsubsection{Dataset}
Though a valuable and unique dataset, the U.S. House of Representatives Permanent Select Committee on Intelligence does not detail how this representative sample of 3.5K ads was selected and redacted.
DiResta et al.~\cite{DiResta_IRA_report}, who were given access to all 61.5K Facebook ads, allude to the bias inherent in the dataset: the social media platforms did not report a methodology, did not include anonymized user comments, and gave minimal metadata.
Furthermore, because the dataset was made available as PDF documents, we leveraged the PyPDF2 Python library to automate text extraction from the files. 
As such, some parsing errors (e.g., \texttt{8} being confused as \texttt{0}) could have been present in our data and may have slightly affected our data collection. 
This emphasizes the need for rich and diverse ground-truth disinformation datasets for future research on the topic.

\subsubsection{Data Analysis}
We are aware that we limited our correlation analyses to pairwise correlations, which undoubtedly do not fully summarize the relationship between our features. 
Additionally, works such as Aldous et al.~\cite{aldous2019view} detail the many levels and metrics of engagement.
Our dataset contained only the metadata for impressions and clicks, and although these two metrics were highly correlational, our analysis may have benefited from a ``composite'' engagement score that leverages both features, instead of considering them separately. 
Future work should aim to compare the distribution of features for High and Standard Engagement groups instead of relying solely on mean values to infer the polarity of Maximal Correlation results.

Our data collection extracted the ad texts but discarded any images associated with the ad, overlooking the presence of emotionally charged visual stimuli used in combination or as its own malicious ad product (see Fig.~\ref{fig:visceral_images} for examples). 
To mitigate this data loss, future works can leverage deep learning architectures such as neural networks for image captioning to characterize the content of an image~\cite{Hossain2018-dd} and pair it with the ad text and dataset's features. 
Similarly, if an ad only contains a video, future works can make use of video summarization and image captioning with attention-based mechanisms~\cite{Fajtl2018-la} to leverage all the available information. 
In fact, this treatment of media files has its standalone merit and is well suited to be integrated within social media platforms.

In our work, we analyzed a collection of features readily available and extractable from the dataset (such as ad metadata and text size) but we also extracted several other sentiment and subjectivity features using the pre-trained NLTK VADER, TextBlob and Flair models, and sociolinguistic features using the novel text analysis tool LIWC. 
Our analysis includes full characterization of the relationship between these features and the dependent variable (ad clicks, i.e., engagement) including both linear and non-linear relationships. 
However, due to the inherent sparsity and noisiness in natural language processing, the extracted features will quickly become co-linear which can impact subsequent feature selection techniques.
In the future, we plan to use techniques such as the Gram-Schmidt Transform~\cite{Wang2016-ci} to guarantee orthogonality of the feature space. 

We also corroborated prior works by~\cite{DiResta_IRA_report, howard18} showing that the IRA purposefully targeted communities to polarize political discourse in the U.S.
Based on this, the unsupervised LDA performed surprisingly well considering the relatively small dataset.
LDA is a powerful tool for topic modeling, though it suffers from major drawbacks similar to many unsupervised models, including: 
(1) stasis, that is, LDA finds the set of topics for the entire dataset without the ability to track them over time; 
(2) the number of topics needs to be defined \emph{a priori} (in this work, we applied measures of consistency and reproducibility to determine the best number of topics); 
(3) LDA measures keyword contribution based on a Bag of Words (BoW) model, which assumes words are exchangeable, the sentence structure (semantic) is not modeled; 
(4) non-hierarchical modeling, where keywords are shared between topics; and, finally, 
(5) LDA topic distribution does not capture linear correlation relationships between topics, as intuitively disclosed in drawback (4).

Another limitation of this work refers to our sociolinguistic analysis of the ads' text. 
Out of 85 categories available in the LIWC tool, we restricted our analysis to the 17 main categories to reduce sparsity in our feature space, given our relatively small set of ads. 
Because these features exhibited promising results in predicting engagement, the examination of more LIWC categories using larger sets of ads is a potential fruitful direction of future work. 
Moreover, in our topic modeling analysis, we used BoW to convert the ads' text into a vector representation to serve as an input for the LDA model. 
This may also be a limitation of our study since BoW disregards certain properties of the text such as grammar, semantic meaning, and word ordering. 
The use of other word vectorization techniques able to capture semantic meaning and other relevant properties, such as Word2Vec and GloVe, is therefore another research direction.
\\

Understanding what makes for high engagement in propaganda and disinformation ads paves the way for countermeasures in several respects. 
First, future research can evolve the social media labels and potentially expose deceptive cues in posts from suspicious or biased accounts to better inform users. 
This is particularly important when we consider that Russia Active Measures did not stop after the campaign considered in this work and in fact intensified after the election~\cite{Select_committee_on_intelligence2019-fb}. 
For example, in 2018, the Washington Post reported that Russian trolls inflamed the U.S. debate over climate change~\cite{Craig_Timberg2018-jo}. 
In June of 2020, the Associated Press reported that U.S. officials confirmed that Russia was behind the spreading of disinformation about the coronavirus pandemic~\cite{Tucker2020-fd}.
Disinformation campaigns have also been generated from own nation states figures~\cite{Guynn2020-gh}, as we have witnessed in the aftermath of the U.S. 2020 presidential election.
The success of such campaigns has even prompted the business of disinformation-as-a-service, which key stakeholders, including disinformation researchers should pay a closer look~\cite{Shelby_Grossman2020-ou}.

\section{Conclusion}\label{sec:conclusion}

This paper focused on a statistical and multi-methods machine learning investigation of features that predict engagement in a dataset of 3,517 Facebook ads created by the Internet Research Agency (IRA) that serves as ground-truth disinformation and were created between June 2015 and August 2017. 
These ads, made publicly available by the United States House of Representatives Permanent Select Committee on Intelligence, were part of a Russia Active Measures disinformation campaign that sought to influence the U.S. Presidential Election of 2016 and sow division in American society, especially on racial issues. 
We extracted a total of \totalfeat features from this dataset and using correlation analysis, feature selection, and topic modeling, we found that: 
(1) ad expenditure, text size, ad lifetime, and sentiment were recurring important features chosen by six different machine learning models in the makeup of a successful disinformation ad; 
(2) positive sentiment ads were more engaging than negative ads;
(3) sociolinguistic features (e.g., use of religion-related words) were highly important in predicting engagement; and
(4) confirming prior works, the IRA targeted several communities and sociopolitical topics (e.g., reproductive rights, Second Amendment rights) during the 2016 U.S. presidential election cycle. We offer suggestions for future works that may shed light on important aspects regarding the prediction of engagement with disinformation, which we hope can foster the next generation of countermeasures and in-depth analyses.




\bibliography{bib/main}
\bibliographystyle{ACM-Reference-Format}



\end{document}